\documentclass[prb,reprint, twocolumn]{revtex4-1} 

\usepackage{amsmath}
\usepackage{amsfonts} 
\usepackage{graphicx}
\usepackage{hyperref}
\usepackage[T1]{fontenc}
\usepackage{ae,aecompl}
\usepackage{amssymb}
\usepackage{listings} 
\usepackage{float} 
\usepackage{color}

\definecolor{codegreen}{rgb}{0,0.6,0}
\definecolor{codegray}{rgb}{0.5,0.5,0.5}
\definecolor{codepurple}{rgb}{0.58,0,0.82}
\definecolor{backcolour}{rgb}{0.95,0.95,0.92}

\lstdefinestyle{mystyle}{
	backgroundcolor=\color{backcolour},   
	commentstyle=\color{blue},
	keywordstyle=\color{magenta},
	numberstyle=\tiny\color{codegray},
	stringstyle=\color{codepurple},
	basicstyle=\footnotesize,
	breakatwhitespace=false,         
	breaklines=true,                 
	captionpos=b,                    
	keepspaces=true,                 
	numbers=left,                    
	numbersep=5pt,                  
	showspaces=false,                
	showstringspaces=false,
	showtabs=false,                  
	tabsize=2
}

\begin{document}

\title{GalRotpy: an educational tool to understand and parametrize the rotation curve and the gravitational potential of disc-like galaxies}

\author{Andr\'es Granados}
\email{afgranadosc@unal.edu.co} 
\affiliation{Departamento de F\'isica}
\affiliation{Observatorio Astron\'omico Nacional, Universidad Nacional de Colombia, Carrera 30 Calle 45-03, P.A. 111321 Bogot\'a, Colombia}

\author{Daniel Torres}
\email{daatorresba@unal.edu.co}
\affiliation{Observatorio Astron\'omico Nacional, Universidad Nacional de Colombia, Carrera 30 Calle 45-03, P.A. 111321 Bogot\'a, Colombia}

\author{Leonardo Casta\~neda}
\email{lcastanedac@unal.edu.co}
\affiliation{Observatorio Astron\'omico Nacional, Universidad Nacional de Colombia, Carrera 30 Calle 45-03, P.A. 111321 Bogot\'a, Colombia}

\author{Lady Henao}
\email{ljhenaooc@unal.edu.co}
\affiliation{Observatorio Astron\'omico Nacional, Universidad Nacional de Colombia, Carrera 30 Calle 45-03, P.A. 111321 Bogot\'a, Colombia}

\author{Santiago Vanegas}
\email{svanegasp@unal.edu.co}
\affiliation{Observatorio Astron\'omico Nacional, Universidad Nacional de Colombia, Carrera 30 Calle 45-03, P.A. 111321 Bogot\'a, Colombia}

\date{\today}

\begin{abstract}
	\textbf{GalRotpy} is an educational \verb+Python3+-based visual tool, which is useful to undestand how is the contribution of each mass component to the gravitational potential of disc-like galaxies by means of their rotation curve.  Besides, \textbf{GalRotpy} allows the user to perform a parametric fit of a given rotation curve, which relies on a MCMC procedure implemented by using  \verb+emcee+ package. Here the gravitational potential of disc-like galaxies is built from the contribution of a Miyamoto-Nagai potential model for the bulge/core and the thin/thick disc, an exponential disc, together with the NFW (Navarro-Frenk- White) potential or the Burkert (cored density profile) potential for the Dark Matter halo, where each contribution is implemented by using \verb+galpy+ package. We summarize the properties of each contribution to the rotation curve involved, and then describe how \textbf{GalRotpy} is implemented along with its capabilities. Finally we present the characterization of two galaxies, NGC6361 and M33, and show that the results for M33 provided by \textbf{GalRotpy} are consistent with those found in the literature.
\end{abstract}

\maketitle 

\section{Introduction} 

In 1914 Vesto Slipher discovered that spiral galaxies rotate, by detecting inclined absorption lines in nuclear spectra from M31 and Sombrero galaxies \citep{slipher1914lowell}. Later, Jan Oort in 1932, first found that there must be three times as much mass as it is observed in visible light when he studied stellar motions above the galactic plane. This finding prompted him to include undetected components like interstellar medium to explain the missing mass. Similar Fobservations for the external parts of NGC3115 galaxy, showed that the mass-to-light ratio is about two orders of magnitude larger than in the solar neighborhood \citep{oort1940some}, as evidence of no visible matter. The latter is known as the \textit{missing mass} problem; it is, the mass contained in the bright objects of a defined region in space does not correspond to its dynamical mass, brought to us by its gravitational interactions.\footnote[1]{asdfasdfasdf}

The mass-to-light ratio $\Upsilon = M/L$ is a quantity that describes how much the mass is a fraction of the light expressed in solar units ($\Upsilon_\odot = \textrm{M}_\odot\textrm/{L}_\odot$). It has been the main tool for investigating the \textit{missing mass} problem in stellar systems like the Milky Way galaxy, external galaxies, and cluster of galaxies.

Is has been raised some explanations concerning the \textit{missing mass} problem. H. Babcock in 1939 found that the rotation curve is approximately flat on the periphery of M31 galaxy, instead of the expected Keplerian decrease because of the diminishing in luminosity (predicted by the luminous profile) \cite{babcock1939rotation}. He concluded that the mass-to-light ratio must be not constant in the galactic radius, but it must increase. He suggested two explanations for this phenomenon: the light absorption must increase in external parts of the galaxy, or it is required a modification to the Newtonian dynamics \cite{sanders2010dark}. 

The findings published by Fritz Zwicky \cite{zwicky1933, zwicky1937masses}, suggested the existence of some sort of \textit{unseen matter} or dark matter in his results using the virial theorem applied to the velocities of galaxies in the Coma galaxy cluster. Zwicky measured the radial velocities of the galaxies in the cluster, and thus he estimated the cluster mass as well as the average galaxy mass. Then comparing this value with the luminosity, he obtained the mass-to-light ratio for galaxies in the cluster $\Upsilon = 500 \Upsilon_\odot$ suggesting that the major contribution comes from dark matter in the cluster. Later, in 1970 Vera Rubin, first reveals an observational evidence of dark matter in M31 galaxy. She realized the flattened circular velocity in the external regions of the galaxy based on the galaxy rotation curve from 67 H\textsc{ii} spectra within a range in the galactic radius of ($3-24kpc$) \cite{rubin1970rotation}.

The rotation curve of disc-like galaxies is the main kinematic observable data that allow the study of dynamical properties of its stars and interstellar gas, in addition to structure, evolutionary and formation processes of the galaxy \cite{sofue2001rotation}. The shape of rotation curves was related to the morphology of spiral galaxies \cite{rubin1980rotational} looking for a universal rotation curve depending only on the galaxy luminosity \cite{persic1996universal}, and not only on the luminosity but by a multi-parameter family such as morphological type, the shape of the light distribution and other optical properties \cite{noordermeer2007mass}.

The mass distribution in a component of a galaxy can be estimated by the assumption that the mass-to-light ratio is constant \cite{binney2011galactic}. Given that the galaxy luminosity is an astrophysical observable, it can be obtained a light profile of a galaxy component and, therefore to infer its mass distribution. Then, it is possible to find the rotation curve for each mass contribution and derive interesting quantities like bulge to disc or bulge to the dark matter halo mass ratios, or equally interesting the radial extension of each mass component. The modeling of a rotation curve using mass decomposition is widely used even in recent studies, since it evidences the influence of all the mass contributions in each position.

\begin{figure}
	\centering
	\includegraphics[scale=0.59]{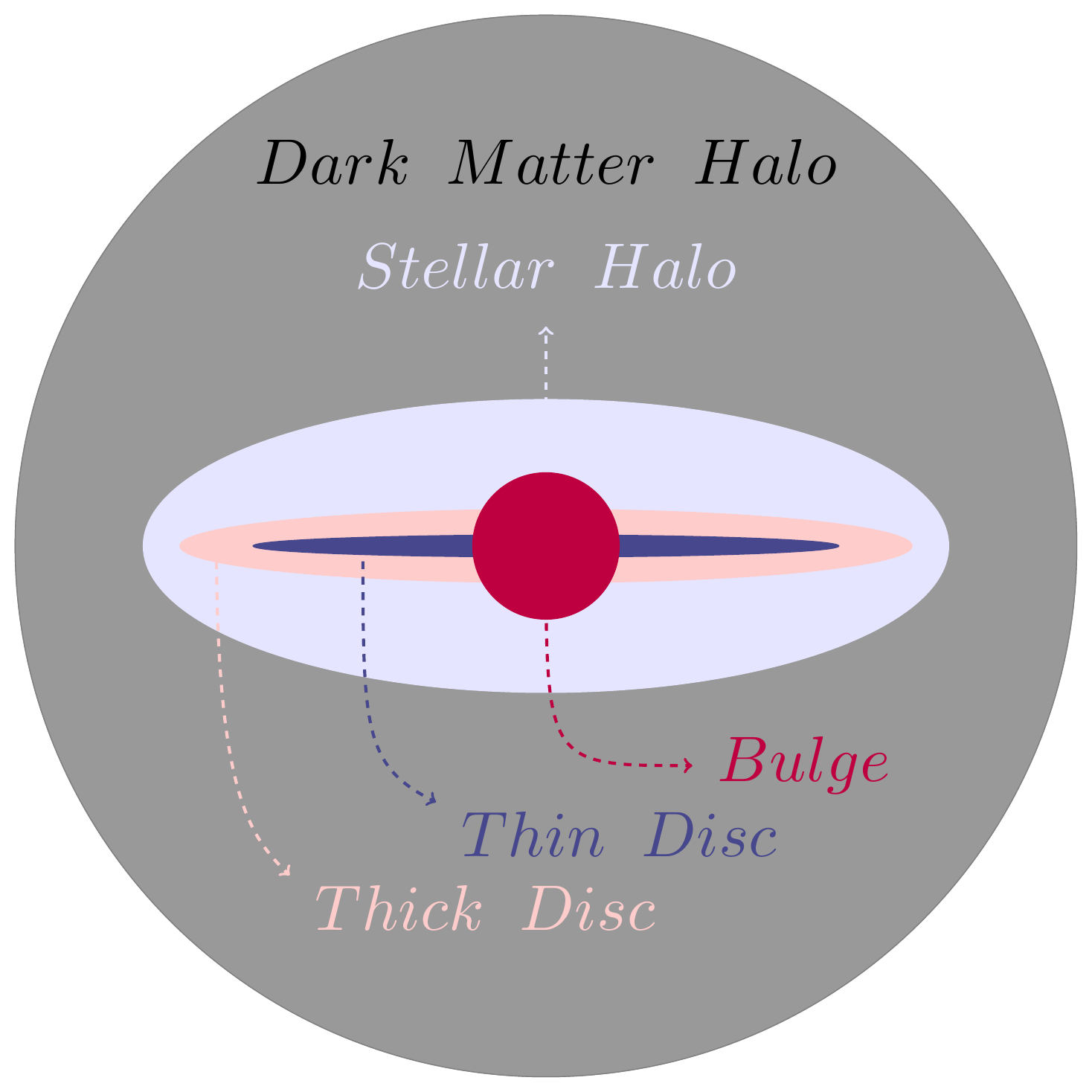}
	\caption{ Diagram of the main components of disc-like galaxies: spheroidal bulge, thin and thick discs, spheroidal stellar and Dark Matter halos.\\}
	\label{fig:Fig_diagram_galaxy}
\end{figure}

Following the method of mass decomposition, \textbf{GalRotpy} is intended to visualize the parametric building of a rotation curve of disc-like galaxies in its main mass constituents and then to make an estimation of the parameters and their uncertainties associated to each contribution. A rotation curve can be fitted by changing in real time two or more gravitational potential parameters, to give a first glance about the importance of the different mass components of a galaxy, that can be compared directly with observational data. Therefore \textbf{GalRotpy} is also presented as a teaching guide for gravitational potential theory and dynamics of galaxies.

This paper is arranged in the following way. In section \ref{sec: gravitational potentials} we summarize the theory of gravitational potential as the foundations of \textbf{GalRotpy}, where we focus on the functional form of four gravitational potentials used to model the galaxy's mass components and their circular velocities: Miyamoto-Nagai for the bulge and thin/thick disc, the exponential disc and the Navarro-Frenk-White or Burkert for the dark matter halo. Here, we also make a brief discussion on how this decomposition has been implemented\cite{sofue1996, sofue2016rotation, pouliasis2017milky}, which is a reference for defining the input parameters used in \textbf{GalRotpy}. 
This is followed by section \ref{sec:galrotpy}, in which we give the outline of how \textbf{GalRotpy} is implemented and how it works. Continuing with section \ref{sec:results}, it is presented the application of \textbf{GalRotpy} to two spiral galaxies, NGC6361 and M33, in order to estimate the potential parameters like mass (or density), galactocentric distance and height scales from the rotation curve. Finally, the conclusions and summary are subject of section \ref{sec:conclusions}. 

\section{Gravitational potential of disc-like galaxies}\label{sec: gravitational potentials}

In this section we take into account the primary results on the theory of gravitational potential related to different mass components in disc-like galaxies; we also show the equations of circular velocity and the meaning of the parameters for these potentials to understand the basis to decompose the observed rotation curve of disc galaxies. The potential theory is the fundamental issue needed to extract the kinematical features from rotation curves, and deduce the dynamics of these systems.

\subsection{Potential Theory}

A galaxy is a system of stars, interstellar gas and dark matter that interact between them fundamentally following Newton's theory of gravity. The whole mass of a disc-like galaxy is composed of different masses associated with its constituents stellar systems. The mass distributions will give us the functional form for the potentials according to the Newton gravity law expressed in its differential form by \textbf{Poisson's equation} \cite{samurovic2007dark, binney2011galactic}:
\begin{equation}
\label{PT2}
\nabla^2 \Phi (\textbf{x}) = 4\pi  G \rho (\textbf{x}),
\end{equation}
with  $G$ the gravitational constant, $\rho$ the mass density of the given system and $\Phi (\textbf{x})$ the gravitational potential this systems generates. From now on we will refer to $\Phi=\Phi (\textbf{x})$ as potential.
 
The Poisson's equation is an elliptical partial differential equation that allows linearity. It means  that if two sources with mass density $\rho_1$ and $\rho_2$ generate the potentials $\Phi_1$, and $\Phi_2$ respectively, then the source with mass density  $\rho=\rho_1 + \rho_2$ generates the potential $\Phi=\Phi_1+ \Phi_2$. Here linearity  is also known as \textit{the superposition principle }, and it stays valid for as many mass components as needed. This is the key property implement in \textbf{GalRotpy} tool,  since it allows the decomposition of the total potential of a given galaxy into its different components, and thus use them separately.  

When we can set $\Phi(\textbf{x}\rightarrow \infty) = 0$ the solution to (\ref{PT2}) is
\begin{equation}
\label{PT1}
\Phi(\textbf{x}) = -G \int d^3   \textbf{x}^{\prime} \frac{\rho(\textbf{x}^{\prime})}{| \textbf{x}-\textbf{x}^{\prime} |},
\end{equation}
where the integral is taken over all the mass distribution, such that (\ref{PT1}) satisfies $\textbf{F}=-\nabla \Phi$, with $\textbf{F}$ being the total force per unit mass on a particle, also known as the gravitational field.

For a deep discussion on potential theory in the gravitational context refer to \citet{binney2011galactic}.

\subsection{Circular velocity}

The central issue of this work is to compute the circular velocity $V_c$ (in the equatorial plane) associated to a gravitational potential $\Phi (R,z=0)$ of a disc-like galaxy, which is described by
\begin{equation}
\label{CV1}
V_{c}^2(R) = R\frac{\partial \Phi}{\partial r}\bigg\vert_{r=R},
\end{equation}
where in case of a spherical symmetric distribution, this velocity is not fixed at the equatorial plane, and reads
\begin{equation}
\label{CV3}
V_{c}^2 = \frac{G}{R}M(\leq R),
\end{equation}

Thus, along to the linearity of Poisson's equation (\ref{PT2}), we have that, for a system  composed by $n$ mass distributions, it is characterized by the gravitational potential 
\begin{equation}
\label{CV2}
\Phi = \Phi_T = \sum_i^n \Phi_i,
\end{equation}
such that, according to equation (\ref{CV1}) and the potentials composition (\ref{CV2}), the total circular velocity reads
\begin{equation}
\label{CV3}
V_{c}^2 = \sum_i^n V_{c(i)}^2.
\end{equation}

Therefore, using equation (\ref{CV1}), we obtain the circular velocities associated to each potential that we will use to model the rotation curve of a disc-like galaxy \cite{binney2011galactic}.

\subsection{Potentials of a disc-like galaxy}\label{sec: potentials}

It is a tough task to resolve the Poisson equation for $\Phi(\textbf{x})$ given a mass density $\rho (\textbf{x})$. However, supported by symmetry considerations and observed luminosity profiles it is possible to simplify the problem and to find a functional form for the mass distribution of a spiral galaxy\citep{miyamoto1975three}.

There are different combinations of potentials which permit to characterize the rotation curve of a given disc galaxy, for example in \citet{sofue1996} the rotation curve for the Milky Way galaxy was modeled using the Miyamoto-Nagai potential for four different componets, while in \citet{pouliasis2017milky} for this very same rotation curve, two models are given, where Miyamoto-Nagai potential, Plummer potential and a truncated spherical symmetric potential are used to model the thin/thick disc, the bulge and the dark halo respectively. In this way,  \citet{sofue2016rotation} modeled the rotation curve of several disc galaxies by means of the de Vaucouleurs potential for the bulge, an exponential disc and a NFW dark halo. Besides, it has been shown that  Burkert cored distribution is useful to describe the dark halo for dwarf galaxies for example (see \citet{karukes2016universal}). 

Therefore, we have a guide of what potentials (mass distributions) are the best to be include in \textbf{GalRotpy}, taking into account that we will use \verb+galpy+ to implement the different contributions of a given rotation curve. Thus, as a first approximation to the real dynamics of disc-like galaxies, \textbf{GalRotpy} uses  calculations of two axisymmetric potentials to model the barionic matter component, while it uses two spherical symmetric potentials to model the dark matter component. The mass distribution of a disc galaxy can be decomposed mainly into three mass components: bulge, disc, and Dark Matter Halo \citep{sofue2016rotation}. Below, we present in context the historical development of these components, the functional forms for its gravitational potentials and its importance in the understanding of the rotation curve of disc-like galaxies. 

\subsubsection{Miyamoto-Nagai potential}

This potential expresses the mass components of the bulge and the thin/thick disc of a galaxy. The Miyamoto - Nagai potential is a generalization of the Plummer and Kuzmin potentials.

In 1911, Plummer used an elementary solution of the Lane-Emden equation to find the gravitational potential of a spherical system (refer to \citet{binney2011galactic} for further details), which is know as the Plummer model, and it is given by
\[
\Phi_P (R,z) = - \frac{GM}{\sqrt{R^2+z^2+b^2} } = - \frac{GM}{\sqrt{r^2+b^2} }
\]
with $r^2 = R^2+z^2$. The Plummer model raised on the discussion about stars distribution in globular clusters. The mass density at different distances from the center of globular clusters is approximately the same, and it is supposed these systems come from spherical distributions \citep{plummer1911problem}.

Moreover, for a flattened mass distribution of axisymmetric galaxies, Toomre in 1963 found the exact solutions for the Poisson's equation
\[
\nabla^2 \Phi (\textbf{x}) = 4\pi  G \mu(R) \delta(z),
\]
where $\mu(R)$ and $\delta(z) $ are the surface density and Dirac's delta function respectively. Its solution \cite{miyamoto1975three, binney2011galactic} turns out to be 
\[
\Phi_K (R,z) = - \frac{GM}{\sqrt{R^2+(a+|z|)^2} },
\]
which is named the Kuzmin's model or Toomre's model 1.

On the other hand, the Miyamoto - Nagai potential is an axisymmetric potential defined in cylindrical coordinates $(R,z)$ as
\begin{equation}
\label{MN1}
\Phi_{MN} (R,z) = - \frac{GM}{\sqrt{R^2+(a+\sqrt{z^2+b^2})^2}},
\end{equation}
with $a$, $b$, $M$ the length, height scales and mass enclosed at galactocentric distance $R$ (amplitude), respectively; these can be used as free parameters to determinate the mass components. Thus, for this potential its associated circular velocity reads 
\begin{equation}
V_c (R) = R \sqrt{ \frac{GM}{ (R^2+(a+b)^2)^{3/2}}  }
\end{equation} 

\begin{figure}
	\centering
	\includegraphics[scale=0.58]{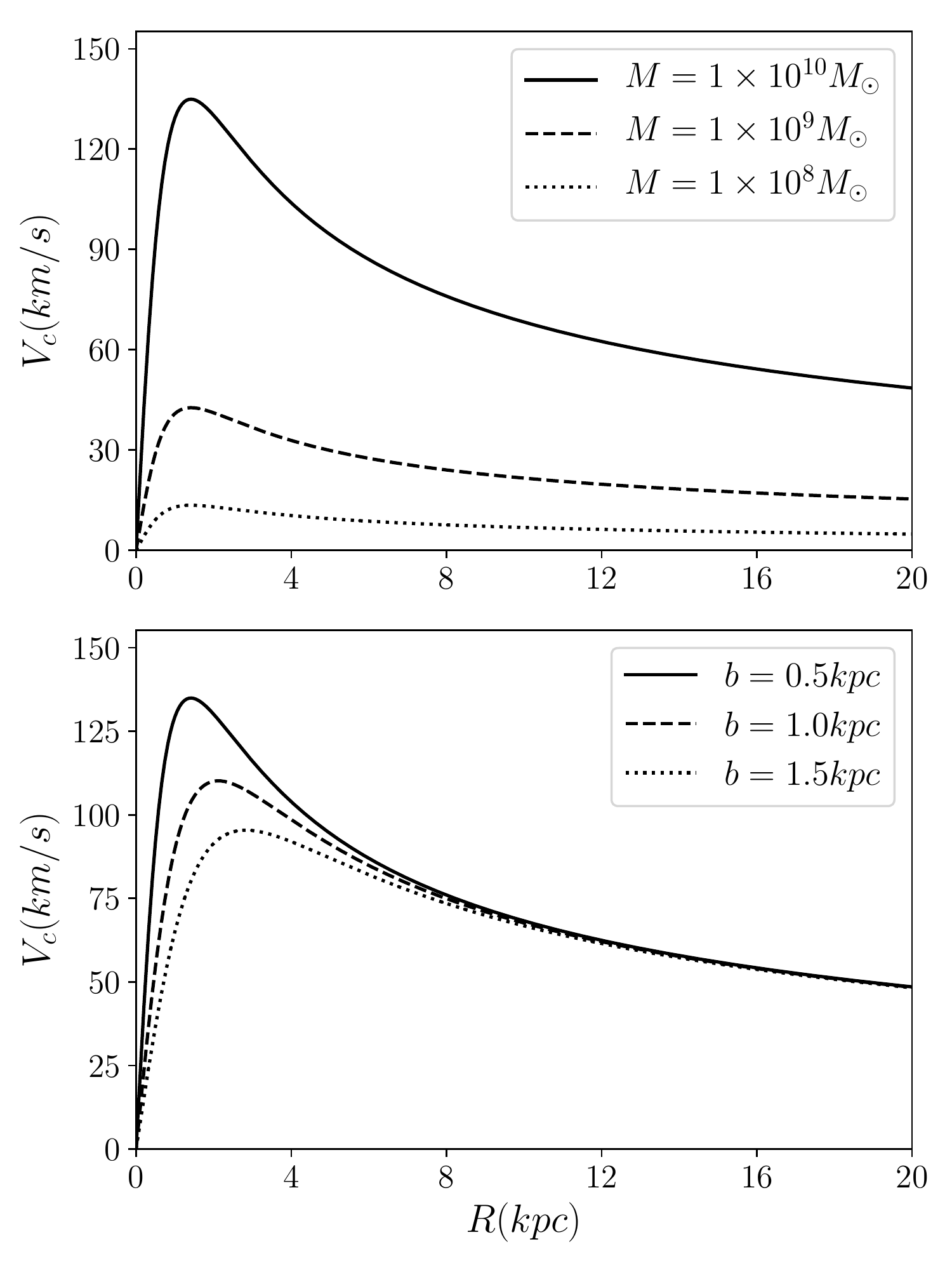}
	\caption{ Rotation curve for the Miyamoto-Nagai potential with three values of the amplitude $M$ (top) and the scale parameter $b$ (bottom).\\}
	\label{fig:Fig_MN_parameters}
\end{figure}

The Fig. \ref{fig:Fig_MN_parameters} represents the changes in the rotation curve's shape by three values of amplitude. These values are traced by the quantity $M$ (top) representing the mass in the Miyamoto-Nagai model and the height scale $b$ (bottom) that determines the flatness of the mass density (better traced by the dimensionless scale ratio $b/a$). The model \ref{MN1} is free of singularities and tends to the Newtonian point mass potential when $R$ and $z$ become large \citep{miyamoto1975three}.

This potential is a generalization of the Plummer and Kuzmin potentials $\Phi_P (R,z)$ and $\Phi_K (R,z)$ respectively. A spherically symmetric potential (i.e. a Plummer model) can be expressed using the scale parameters defined in the Miyamoto-Nagai potential (\ref{MN1}) with $a=0$ \citep{samurovic2007dark}, while the Kuzmin model is recovered when $b=0$.

The Miyamoto-Nagai potential describes both a disc and a bulge geometries continuously, including Plumer and Kuzmin potentials without superposing them \citep{miyamoto1975three}. This potential is implemented in \verb+galpy+ and is used to model the bulge, thin disc and thick disc when needed to reproduce the rotation curve of a given disc-like galaxy.

\subsubsection{Razor thin exponential disc potential}

An axisymmetric thin disc can be considered as very flattened spheroid. Then in the following, we show the gravitational potential of an infinitely thin spheroid assuming some key physical and geometric conditions.

\begin{figure}
	\centering
	\includegraphics[width=8.5cm,height=7cm]{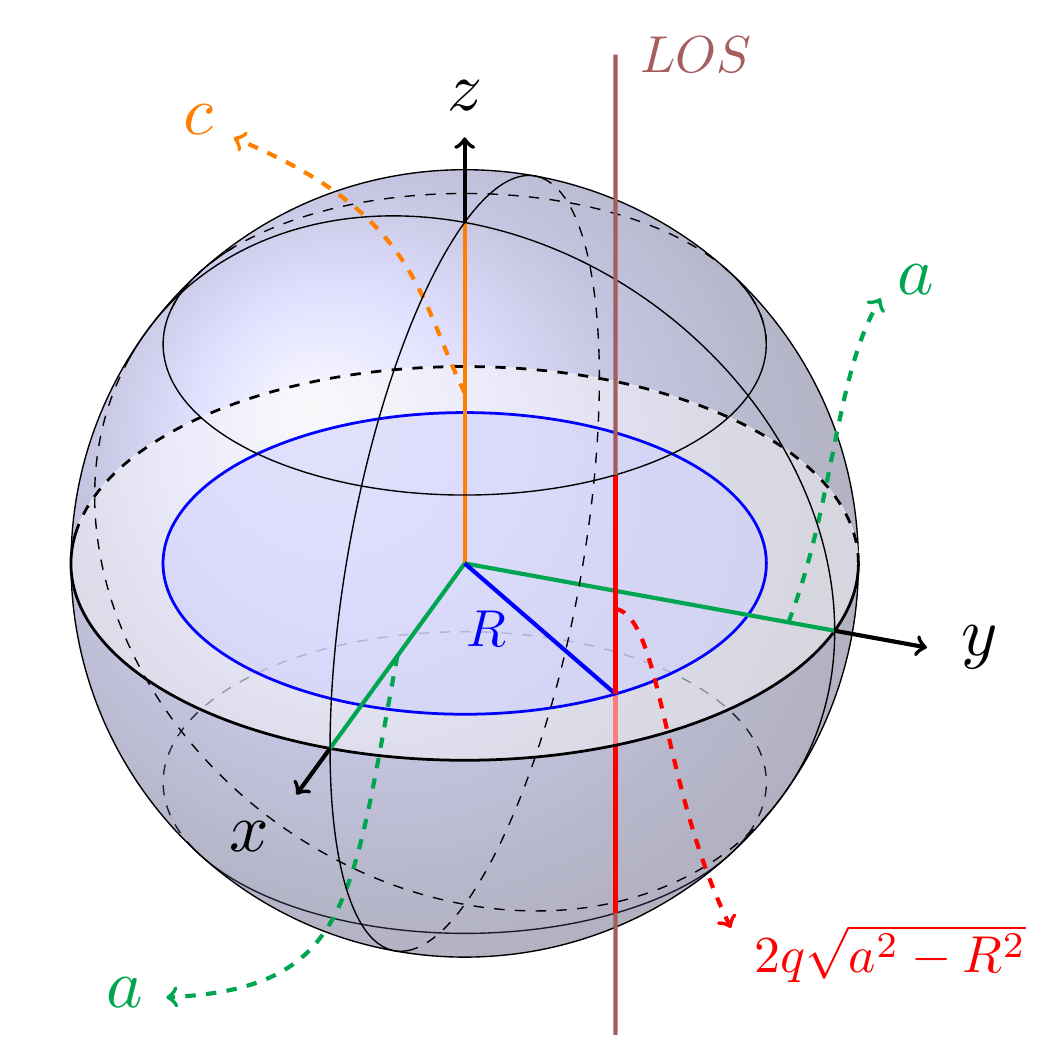}
	\caption{ Spheroid of axis ratio $q=c/a$. The Line Of Sight (LOS) is perpendicular to the equatorial plane and crosses a distance $2q\sqrt{a^2-R^2}$ at a galactocentric radius $R$ from the centre.\\}
	\label{fig:Fig_Galaxy_Spheroid}
\end{figure}

Since the idea is to obtain the potential of a flat spheroid, it is important to describe its source. According to Fig. \ref{fig:Fig_Galaxy_Spheroid} we see that the distance along the polar axis (along the $z$ direction) allows us to express the enclosed mass at $R$; which is defined on the equatorial plane. Thus, the galactic disc can be found by bringing the height of the mass distribution $dz$ towards the equatorial plane. This projection permits to describe the mass content by means of an effective surface mass density, which is commonly noted as $\Sigma$\cite{binney2011galactic}.

The projection described above is a key step in this construction. For such projection it is needed to write the polar coordinate $z$ in terms of the other geometric quantities involved, thus, for a spheroid with axis $a$ and $c$ which satisfy the relation 
\[
\frac{x^2}{a^2} + \frac{y^2}{a^2} + \frac{z^2}{c^2} = \frac{R^2}{a^2} + \frac{z^2}{c^2} = 1,
\]
it is clear that the polar distance can be written as $ z = q\sqrt{a^2 - R^2} $ for an axis ratio $q=c/a$. Therefore, for a homogeneous spheroid with mass density $\rho$ and total mass $M=4\pi qa^3/3$, the surface density is given by $\Sigma (a,R) = \rho D = 2q \sqrt{a^2-R^2}\rho$. Here $D$ is the distance along the Line Of Sight (LOS) which crosses the spheroid (see Fig. \ref{fig:Fig_Galaxy_Spheroid}).

Then, calculating the mass and surface density differentials by a variation in the distance $a$, it is obtained the respective flattened homoeoid quantities
\[
\delta M(a) = 2\pi \Sigma_0 a \delta a \textrm{  ;  } \delta \Sigma(a,R) = \frac{\Sigma_0 \delta a}{\sqrt{a^2-R^2}},
\]
with the central density $\Sigma_0 = 2\rho q a$ at fixed $a$. So, letting the galactocentric radius $R$ as the only parameter for the surface density of a system with axis ratio $q\to 0$ (i.e. $a\to \infty$) this parameter leads to the expression \citep{binney2011galactic}
\begin{equation}
\label{ED4_5}
\Sigma (R) = \int_R^\infty da \frac{\Sigma_0(a)}{\sqrt{a^2-R^2}}. \end{equation}
Here, $\Sigma(R)$ may be considered as an arbitrary observed surface density model. The above equation has the solution given by the Abel integral equation
\begin{equation}
\label{ED5}
\Sigma_0(a) = -\frac{2}{\pi} \frac{d}{da} \int_a^{\infty} dR \frac{R\Sigma (R)}{\sqrt{R^2-a^2}}.
\end{equation}

Now, the next step is concerned about the potential, such that following \citet{cuddeford1993potentials} (and the references therein) it is possible to obtain a particular $\Sigma, \Phi$ pair, given by
\begin{align*}
\Sigma_k (R) =& \frac{k}{2\pi} J_0 (kR) \\
\Phi_k (R,z) =& -G J_0(kR) exp(-k|z| ),
\end{align*}
for $J_0$ a cylindrical Bessel function and $k$ some free parameter. The resulting disc potential turns out to be a superposition of the above components (for more details see \citet{toomre1963distribution}),
\begin{align*}
\nonumber
\Phi (R,z) =& -2\pi G \int_0^{\infty} \int_0^{\infty} f_1\left(R^{\prime},k, R, z\right) dR' dk,
\end{align*}
with
\[
f_1\left(R^{\prime},k, R, z\right):=\Sigma(R') J_0(kR') R' J_0 (kR) \exp{(-|z|k)}.
\]

Using the properties in the integral form of the Bessel function $J_0$, it can be constructed a razor-thin disc potential like an infinitely flattened homoeoid
\[
\delta\Phi= -2{\pi}G\Sigma_0(a){sin^{-1}} \left (\frac{2a}{\sqrt{+}+{\sqrt{-}}} \right )\delta a,
\]
where the notation $\sqrt{\pm} = \sqrt{z^2+(a\pm R)^2}$ is being used. Then the potential for axisymmetric discs with surface density $\Sigma(R)$, decomposed in homoeoids using the equations (\ref{ED4_5}, \ref{ED5}) can be expressed by
\[
\Phi(R,z) = 4G \int_{0}^{\infty} f_2(R,z;a)da 
\]
with
\[
f_2(R,z;a):=sin^{-1} \left ( \frac{2a}{\sqrt{+}+\sqrt{-}} \right ) \frac{d}{da} \int_a^{\infty}  \frac{R'\Sigma (R')dR'}{\sqrt{R'^2 - a^2}}.
\]





Finally, integrating it in $a$, the  potential for an axisymmetric disc in the plane (when $z \rightarrow 0$) is
\begin{equation}
\label{ED2}
\Phi(R,0) = -4G \int_0^{\infty} \frac{da}{\sqrt{R^2-a^2}} \frac{d}{da} \int_a^{\infty} \frac{R'\Sigma (R')dR'}{\sqrt{R'^2 - a^2}}.
\end{equation}

\begin{figure}
	\centering
	\includegraphics[scale=0.58]{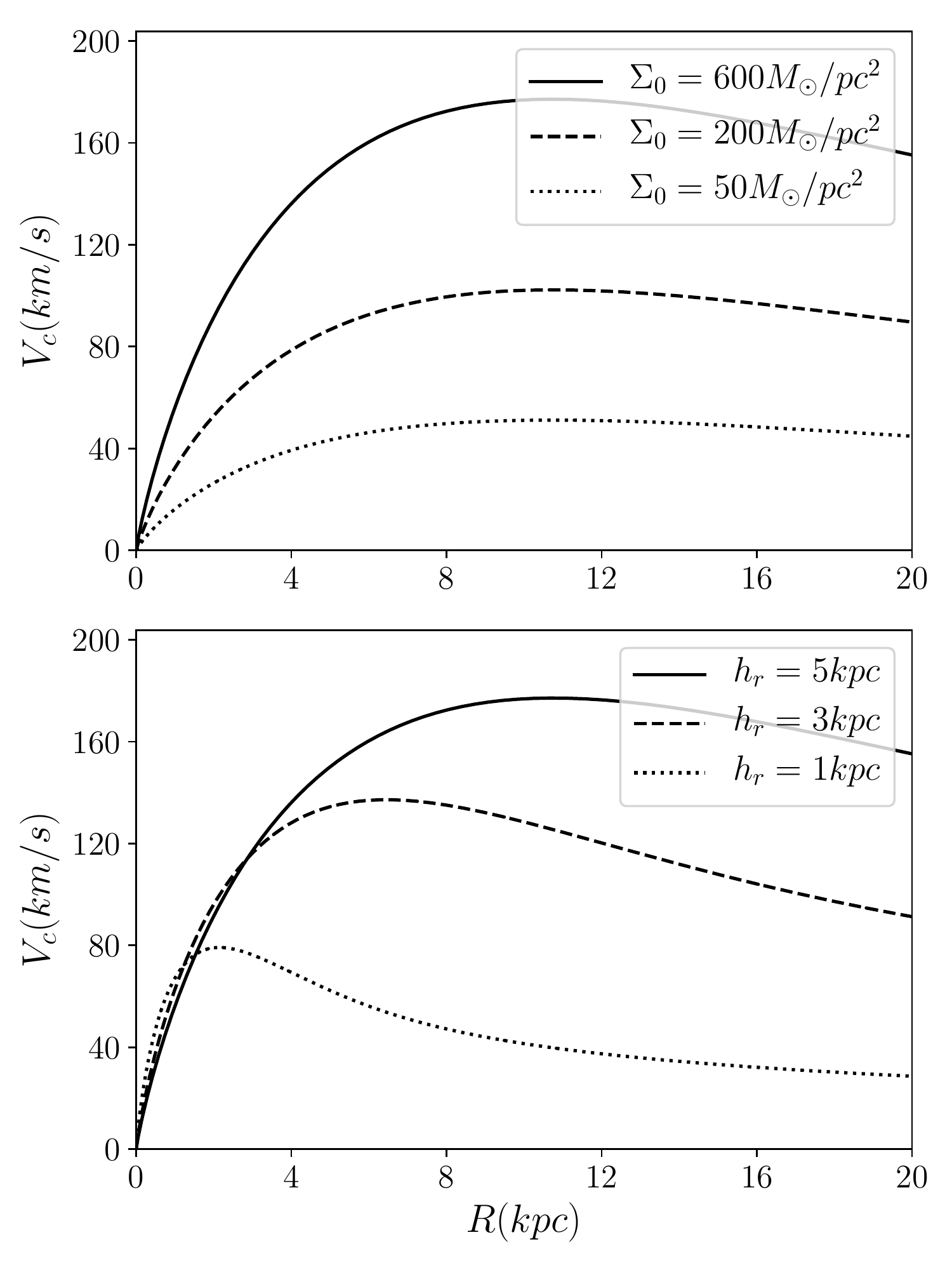}
	\caption{ Rotation curve for an exponential disc potential with three values of the central surface density $\Sigma_0$ (top) and the scale parameter $h_r$ (bottom).\\}
	\label{fig:Fig_ED_parameters}
\end{figure}

According to \citet{freeman1970disks}, it can be assumed that mass-to-light ratio is approximately uniform, at least on the disc of a galaxy and then, the disc surface density profile reads
\begin{equation}
\label{ED1}
\Sigma_d (R) = \Sigma_{0} \exp{(-R/h_r)},
\end{equation}
known as the \textbf{exponential disc}, where $\Sigma_{0}$ and $h_r$ are the central surface mass density and the radial scale respectively. The Fig. \ref{fig:Fig_ED_parameters} shows the change in rotation curve shape by the changes in the potential parameters of equation \ref{ED1}. A change in surface density $\Sigma_{0}$ represents from a very flat circular velocity up to a thick disc (top), and lower values in radius scale parameter $h_r$ allows the reproduction of a thin disc.

Replacing the surface density (\ref{ED1}) in the second integral of the equation (\ref{ED2}), it is obtained:
\[
\int_a^{\infty} \frac{R' \Sigma_0 \exp{\left(-R'/h_r\right)}}{\sqrt{R'^2 - a^2} } dR' = \Sigma_0 a K_1 (a/h_r)
\]
moreover, the potential in the equatorial plane in terms of the modified Bessel functions $K_0, K_1$ and $I_0, I_1$, takes the form
\begin{equation}
\label{ED3}
\Phi_{ED}(R,0) = -\pi G\Sigma_0 R \left ( I_0(y)K_1(y) - I_1(y)K_0(y) \right ),
\end{equation}
which leads to the circular velocity
\begin{equation}
V_c (R) = \sqrt{ 4\pi G \Sigma_0 h_r y^2 (I_0(y)K_0(y) - I_1(y)K_1(y)) },
\end{equation}
where $y=R/2h_r$. For further discussion see \citet{freeman1970disks, binney2011galactic, cuddeford1993potentials}.
For this model the total disc mass is given by \citet{freeman1970disks}:
\begin{equation}
\label{eq:mass exp disc}
M_d = 2\pi h_{r}^{2} \Sigma_{0},
\end{equation}
which depends only on the two free parameters, the scale radius $h_r$ and the central surface density $\Sigma_{0}$.

A complementary method is given by the definition of the galaxy's luminosity by unit of area or \textit{surface brightness} \citep{binney2011galactic}:
\[
I = \int_0^{\infty} dr j(\bf{r}) 
\]
with $ j(\bf{r}) $ the distribution of luminosity density. Given the notation of Fig. \ref{fig:Fig_Galaxy_Spheroid}  in cylindrical coordinates the projection of a spherical body on the $xy-$plane where the LOS is along the $z$ coordinate, results in the surface brightness:
\[
I(R) = 2\int_0^{\infty} dz j(\bf{r})
\]
so that, taking into account the geometric constraint $ r^2 = R^2 + z^2 $ it is clear that $ z = \sqrt{r^2-R^2} $, thus $dz = r dr / z$. The above equation takes the form
\[
I(R) = 2 \int_R^{\infty} \frac{j(\textbf{r}) r dr}{\sqrt{r^2 - R^2}},
\]
which can be inverted in a direct way using the Abel integral identity to obtain $j(\textbf{r})$; that is
\[
j(\textbf{r}) = -\frac{1}{\pi} \int_r^{\infty} \frac{dI}{dR} \frac{dR}{\sqrt{R^2-r^2}}.
\]

Under the assumption of a particular value for the mass-to-light ratio, the volumetric mass density $\rho(r)$ is proportional to the brightness function $j(\textbf{r}) $, that is 
\[
\rho(r) = \frac{M}{L} j(\textbf{r}).
\]

Therefore, starting from the 2D surface density $I(R)$ it can be found the 3D luminosity density $ j(\textbf{r}) $ and if the light traces the mass it can be derived the mass density of the system.\\

\subsubsection{The Navarro-Frenk-White potential} Collisionless N-body numerical simulations of the clustering of dark matter particles suggest that the mass density within a Dark Matter halo has a similar structure to a power density model, and a universal scale behavior. It is interesting to see the similarity between the luminosity profile in elliptical galaxies \citep{binney2011galactic} and the mass distribution in the Dark Matter halo. Such mass density is given by the two-power law
\begin{equation}
\rho (r) = \frac{\rho_0}{(r/a)^\alpha (1+r/a)^{\beta-\alpha}}.
\end{equation}

In particular  for $(\alpha, \beta) = (1,3)$ it is called  Navarro-Frenk-White (NFW) \citep{navarro1997universal} model. This model has two free parameters: the scale radius  $a$ and the representative density $\rho_0$. It has also two correlated parameters: the halo mass $M_a(R)$ and its characteristic (dimensionless) density $\delta_0$ \citep{navarro1997universal}. Finally the NFW model density is
\[
\rho (r) =  \frac{\rho_0}{(r/a) (1+r/a)^2} = \rho_c \frac{\delta_0}{(r/a) (1+r/a)^2},
\]
where $\rho_c = 3H_0^2/8\pi G $ is the cosmological critical mass density, and $\delta_0$ is known as the  characteristic ( dimensionless) overdensity. Here, we use the Hubble parameter $H_0=67.8\pm 0.9$ taken from Planck collaboration \citet{ade2016planck}. 

Thus, the enclosed mass within a radius $r$ is (from \citet{jimenez2003dark})
\begin{equation}
M_{NFW}(\leq r) =M_0 \left[  \ln \left (  1+\frac{r}{a} \right ) - \frac{r/a}{1+r/a} \right ],
\label{eq:NFW_M}
\end{equation}
with $M_0= 4\pi \rho_0 a^3$. This density profile leads to the potential
\begin{equation}
\label{NFW0}
\Phi_{NFW} (r) = -4\pi G \rho_0 a^2 \frac{\ln (1+r/a)}{r/a},
\end{equation}
such that its corresponding circular velocity is
\begin{equation}
V_c (R) = \sqrt{\frac{G}{R}M_{NFW}(\leq R) } 
\end{equation}

\begin{figure}
	\centering
	\includegraphics[scale=0.58]{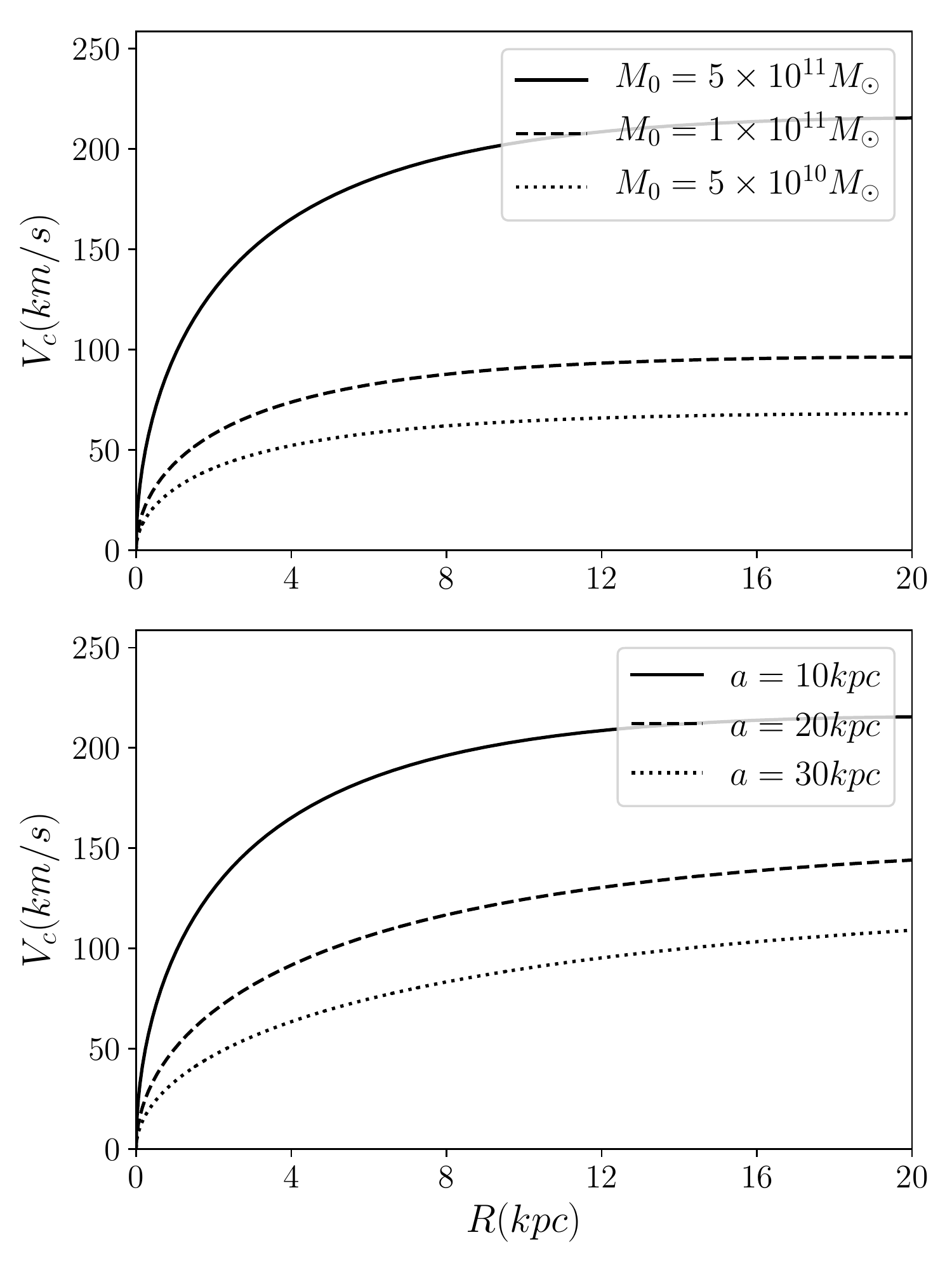}
	\caption{ Rotation curve for the Navarro-Frenk-White potential with three values of the amplitude $M_0$ (top), and the radius scale parameter $a$ (bottom).\\}
	\label{fig:Fig_NFW_parameters}
\end{figure}

As it can be seen in Fig. \ref{fig:Fig_NFW_parameters}, an increment in the quantity $M_0$ results in greater amplitude and any change in scale parameter $a$ gives a fast steep or larger coverage in the galactocentric distance of the galaxy. \textbf{GalRotpy} works directly over the scale $a$ and the effective mass $M_0$ since it is the amplitude parameter implemented in \verb+galpy+.

\subsubsection{The Burkert density profile} 
Some dwarf galaxies are completely dominated by Dark Matter, and the density profiles are according to the density in the center given by the modified isothermal law (\ref{isothermal}). It satisfies the outer rotation curve constant 
value because it falls proportionally to $r^{-2}$:
\begin{equation}
\label{isothermal}
\rho(r) =	\frac{\rho_0}{1+r^2/a^2},
\end{equation}
where $a$ is the core radius and $\rho_0$ is the central dark matter density\citep{burkert1995structure}.

Nevertheless, cosmological simulations in the Cold Dark Matter (CDM) scenario predict halos with central density cusps that are not observed for several dwarfs, spirals and Low Mass Brightness (LMB) galaxies 
\citep{lopez2017radial}. The observed mass profiles of dwarf galaxies can be fitted by the phenomenological density distribution according to \citet{burkert1995structure}, known as the Burkert density profile, which is given by
\begin{equation}
\label{Burkert_profile}
\rho_{Bk}(r) = \frac{\rho_0 a^{3}}{(r+a)(r^2+a^2)},
\end{equation}
such that the mass  enclosed within a radius r is
\begin{align}\nonumber
M_{Bk}(\leq r) = 
&\pi\rho_0 a^3\bigg[  2\ln \left (  1+\frac{r}{a} \right )+\ln\left (  1+\left(\frac{r}{a}\right)^2 \right )\\
& - 2\tan^{-1}\left(\frac{r}{a}\right) \bigg].
\label{eq:burkert_M}
\end{align}
This mass distribution generates a potential given by
\begin{align}
\label{Burkert0}
\nonumber
\Phi_{Bk} (r) =
& \pi G \rho_0 a^2 \bigg\lbrace\left(1-\frac{a}{r}\right)\ln\bigg(1+\left(\frac{r}{a}\right)^2\bigg)\\ 
&2\left(1+\frac{a}{r}\right)\left[\tan^{-1}\left(\frac{r}{a}\right)-\ln\bigg(1+\frac{r}{a}\bigg)\right]\bigg\rbrace,
\end{align}
whose corresponding circular velocity turns out to be
\begin{equation}
V_c (R) = \sqrt{\frac{G}{R}M_{Bk}(\leq R) }.
\end{equation}

\begin{figure}
	\centering
	\includegraphics[scale=0.58]{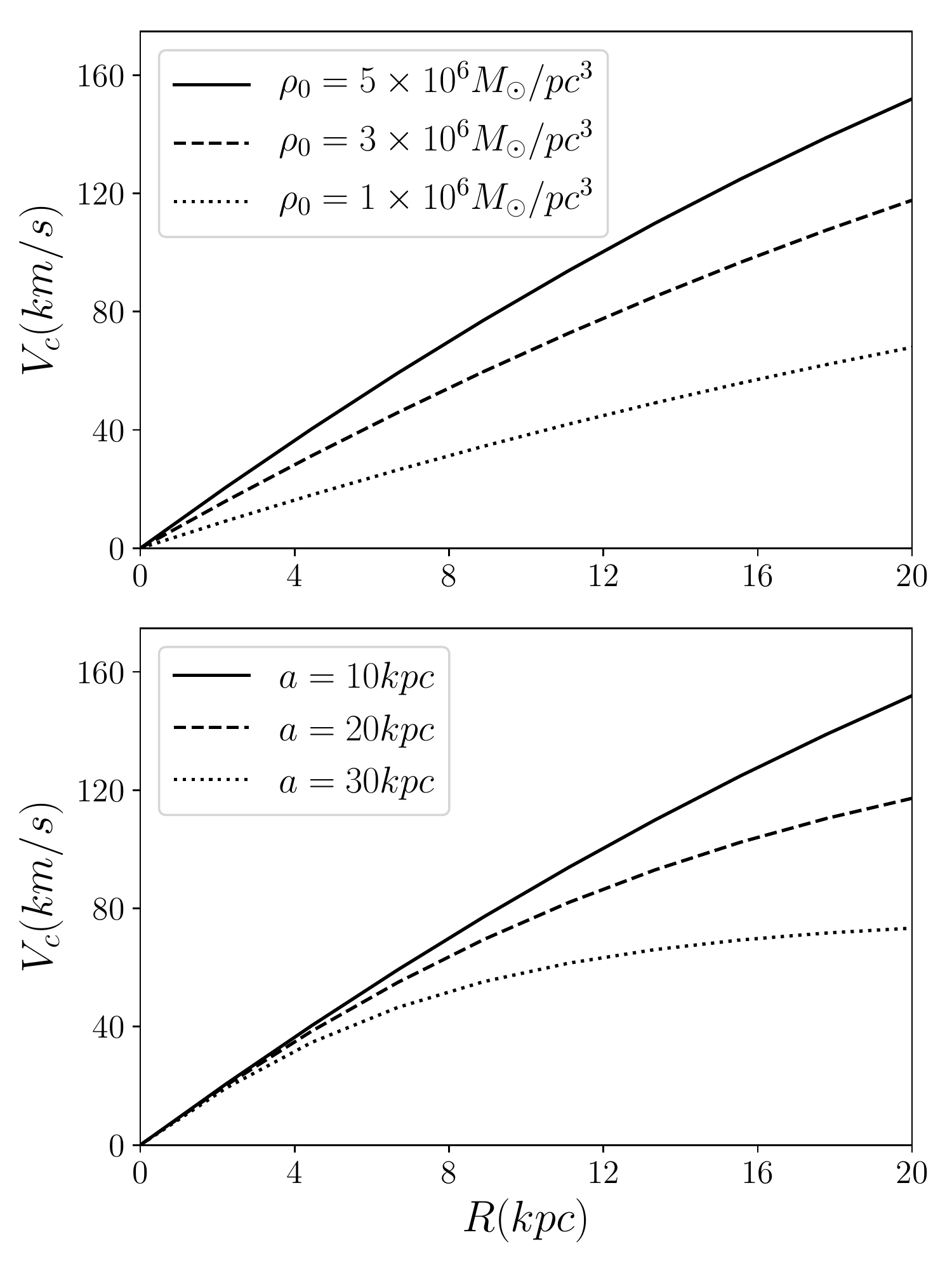}
	\caption{ Rotation curve for the Burkert potential with three values of the central core density $\rho_0$ (top) and the radius scale parameter $a$ (bottom).\\}
	\label{fig:Fig_BK_parameters}
\end{figure}

Here $\rho_0$ and $a$ are parameters which represent the central core density and a scale radius respectively. The corresponding rotation curve is shown in Fig. \ref{fig:Fig_BK_parameters} where a change in the density $\rho_0$ (top) gives a specific value in the amplitude of the cored profile and a lower radius scale value $a$ (bottom) results in a lower top velocity for the same central slope. An interesting property of this profile is that, for practical purposes, it may be characterized by only one of the two parameters described above, since there is an approximate linear relation between $a$ and $\rho_0$ (see \citet{burkert1995structure, salucci2000dark}), which states
\begin{equation}
\label{eq:burkert_linear}
\rho_0\approx 4.5\times 10^{-2}\left(a/kpc\right)^{-2/3}M_{\odot}/pc^3.
\end{equation}

Despite, this seems to be an advantage, we will work with $a$ and $\rho_0$ separately since equation (\ref{eq:burkert_linear}) may induce biased results within the fitting process. Tt means while the walkers explore the parameters space (see section \ref{sec:GalRotpy-Fitting}). With respect to its implementation, unlike NFW profile, for Burkert profile \verb+galpy+ (and therefore also \textbf{GalRotpy}) uses as amplitude parameter $\rho_0$ directly, and $a$ as its scale factor.

\subsubsection{Dark Halo's Mass}\label{sec:DH mass}
The intrinsic parameters of each distribution described above are the best ones to describe a given dark halo, nevertheless in the literature the total mass $M_h$ is often used instead of the corresponding density $\rho_0$, which makes useful to give a brief description of how $M_h$ is defined.

Theoretical work and computational simulations on gravitational collapse and cosmological structure formation have shown that the total mass (also known as critic mass) of a  Dark halo is well defined as the mass enclosed by a limiting radius $r_c$, within which the mean mass density of the halo $\overline{\rho}_h$ is given by
\begin{equation}\label{eq:halo mean density}
	\overline{\rho}_h = \Delta_c\rho_c,
\end{equation}
such that 
\begin{equation}\label{eq:halo total mass}
M_h = M(\leq r_c) = \frac{4\pi}{3}\overline{\rho}_h r_c^3=\frac{4\pi}{3}\Delta_c\rho_c r_c^3.
\end{equation}

Here $\Delta_c$ is the cosmological overdensity and $\rho_c$ is the cosmological critical mass density, which was defined above. The main problem with this definition is that $\Delta_c$ depends on the author and also on the cosmological model used. However, $\Delta_c=200$ is often taken as a standard value. For a deeper discussion about this topic see for example \citet{mo2010galaxy, coe2010dark} and the references therein.

Now, it is clear that in order to obtain $M_h$ we need $r_c$, which is obtained by solving equation (\ref{eq:halo total mass}) for a given profile and, its corresponding intrinsic parameters. In such way, defining the concentration parameter  $X_{\Delta_c}:= r_c/a$  the NFW profile leads to the equation
\begin{equation}
\label{NFW3}
\frac{\Delta_c \rho_c}{3\rho_0} X_{\Delta_c}^3 = \ln(1+X_{\Delta_c}) - \frac{X_{\Delta_c}}{1+X_{\Delta_c}},
\end{equation} 
while Burkert profile leads to the equation
\begin{align}
\label{Burkert critic radius}
\nonumber
\frac{4\Delta_c \rho_c}{3\rho_0} X_{\Delta_c}^3 = 
& 2\ln(1+X_{\Delta_c}) + \ln(1+X_{\Delta_c}^2)\\
& - \tan^{-1}{\left(X_{\Delta_c}\right)}.
\end{align} 

Thus, solving for $X_{\Delta_c}$ we see that for both equations, non trivial real solutions have to be found numerically; plotting the function at both sides of the given equations shows that there is only one non trivial real solution. Remember that the intrinsic parameters $\rho_0$ and $a$ are different for each profile.

\section{GalRotpy}\label{sec:galrotpy}

\textbf{GalRotpy} is a visual tool whose aim is to help to visualize and also to explore the rotation curve of disc-like galaxies considering the contribution of each component independently, not only from a visual inspection but also through a parametric fit analysis. 

In order to accomplish this task  we make an extensive use of the following six main \verb+python+ packages to run \textbf{GalRotpy}: \verb+matplotlib+ to generate the interface and plots, \verb+astropy+ and  \verb+numpy+ for units and data mangement, \verb+galpy+ to construct the rotation curves, \verb+emcee+ to fit the data and obtain the most likely parameters by means of the MCMC procedure, and \verb+corner+ to plot the credible regions obtained from the fit process. Other packages are involved but not extensively used, then for a more detailed description about \textbf{GalRotpy}, see the repository page \cite{GALROTPY} where the source code is available as well as its requirements and the instructions for its use.

\subsection{GalRotpy input}\label{sec:GalRotpy input}

To initialize \textbf{GalRotpy} it is required to give an initial value for the different potential parameters needed to model the bulge, thin disc, thick disc, exponential disc, NFW-halo and burkert-halo, which are introduced by means of a file named \verb+input_params.txt+. This file must contain an initial value for the mass in $M_\odot$ and, for the galactocentric distance and height scales in $kpc$, along with an associated threshold for each component. This threshold should be established for the mass, as an exponent of 10 in relation to its initial value, while for the size scales it must be a percentage of their corresponding initial values.

The following potential parameters (Table \ref{GalRotpy input}) are based in those used to model the Milky Way galaxy and dwarf galaxies, which can be taken by default.
\begin{table}[ht]
	\centering
	\caption{Set of parameters involved in each of the contributions considered in \textbf{GalRotpy}. The given range for each parameter attempts to be a guide about which values may be used to run this program, particularly for dwarf galaxies and Milky way-like galaxies. When our results are far from the values presented in this table, most likely the results are unphysical. \\}
	\label{GalRotpy input}
	\begin{tabular}{lcc}
		\hline
		\multicolumn{1}{c}{
			\textbf{Component}} & 
		\textbf{Parameter Range}   & 
		\textbf{Units}        
		\\ \hline
		Bulge I          & 
		\begin{tabular}[c]{@{}c@{}}
			$a=0$\\ 
			$0.0<b<0.5$\\ 
			$0.1<M<1.0$
		\end{tabular}
		&\begin{tabular}[c]{@{}c@{}}
			$kpc$\\
			$kpc$\\ 
			$10^{10}M_{\odot}$
		\end{tabular}    
		           
		\\ \hline
		Bulge II         &
		\begin{tabular}[c]{@{}c@{}}
			$0.01<a<0.05$\\ 
			$0.5<b< 1.5$\\ 
			$1<M<5$
		\end{tabular} 
		&\begin{tabular}[c]{@{}c@{}}
			$kpc$\\
			$kpc$\\ 
			$10^{10}M_{\odot}$
		\end{tabular}    
		\\ \hline
		Thin Disc         &
		\begin{tabular}[c]{@{}c@{}}
			$1<a<10$\\ 
			$0.1<b< 1.0$\\ 
			$0.5<M<1.5$
		\end{tabular} 
		&\begin{tabular}[c]{@{}c@{}}
			$kpc$\\
			$kpc$\\ 
			$10^{11}M_{\odot}$
		\end{tabular}
		\\ \hline
		Thick Disc         &
		\begin{tabular}[c]{@{}c@{}}
			$1<a<10$\\ 
			$0.1<b< 15.0$\\ 
			$0.5<M<1.5$
		\end{tabular} 
		&\begin{tabular}[c]{@{}c@{}}
			$kpc$\\
			$kpc$\\ 
			$10^{11}M_{\odot}$
		\end{tabular}          
		\\ \hline
		Exponential Disc         &
		\begin{tabular}[c]{@{}c@{}}
			$2<h_r<6$\\ 
			$1<\Sigma_0< 15$
		\end{tabular} 
		&\begin{tabular}[c]{@{}c@{}}
			$kpc$\\ 
			$10^{2}M_{\odot}/pc^2$
		\end{tabular}
		\\ \hline
		NFW - Halo         &
		\begin{tabular}[c]{@{}c@{}}
			$0.1<a<30$\\ 
			$0.1<M_0< 10$
		\end{tabular} 
		&\begin{tabular}[c]{@{}c@{}}
			$kpc$\\ 
			$10^{11}M_{\odot}$
		\end{tabular}
		\\ \hline
		Burkert - Halo         &
		\begin{tabular}[c]{@{}c@{}}
			$2<a<38$\\ 
			$0.1<\rho_0< 10$
		\end{tabular} 
		&\begin{tabular}[c]{@{}c@{}}
			$kpc$\\ 
			$10^{6}M_{\odot}/kpc^3$
		\end{tabular}
		\\ \hline\\
	\end{tabular}
\end{table}

On the other hand, the rotation curve must be introduced through a file named \verb+rot_curve.txt+ containing three columns with units of $kpc$ for the radial coordinate and $km/s$ for the velocity and its uncertainty.

\subsection{GalRotpy panel}

\begin{figure}
	\centering
	\includegraphics[scale=0.73]{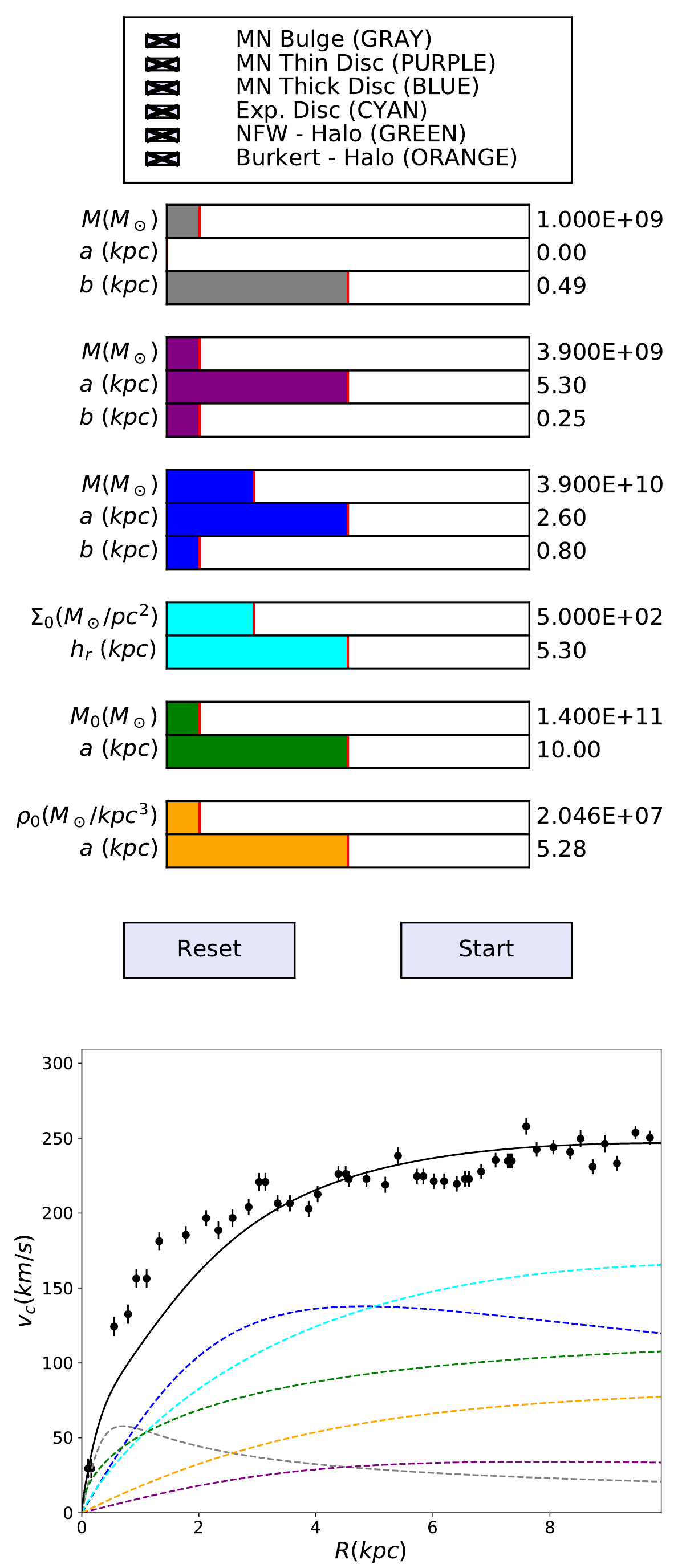}
	\caption{Panel for gravitational potentials selection and parametrization (top) and its composed rotation curve with the corresponding contributions selected (bottom).\\}
	\label{fig:GUI1}
\end{figure}

\textbf{GalRotpy} panel is composed by two blocks, left and right, which are shown in Fig. \ref{fig:GUI1}. First the left block (Fig. \ref{fig:GUI1}---top) includes a checklist to select the potentials to be used to model the rotation curve, and also includes a set of sliders for each mass contribution and scale parameters whose color match their corresponding potential. The sliders present a red guide showing the input parameters described in the previous subsection. Finally at the bottom of this block, there are two buttons: one to reset all the parameters to their input values, and the other one to start the best fit processes. 

Secondly the right block (Fig. \ref{fig:GUI1}---bottom) shows the composed rotation curve given by the black solid line, and also shows the different potentials selected to reproduce the data; each potential is represented by a dashed line whose color matches the corresponding sliders. These rotations curves are obtained by using \verb+galpy+, which is a \verb+python+ library that contains a set of tools for galactic dynamics, including gravitational potentials and its derived quantities: mass density, circular velocity, total mass, among others. \textsc{galpy} performs numerical orbit integration with a variety of Runge--Kutta--type and symplectic integrators; and it supports the calculation of action-angle coordinates and orbital frequencies for spherical potentials. It includes some distribution functions (DF) also like two-dimensional axisymmetric and non-axisymmetric disc DFs, a three-dimensional disc DF, and a DF framework for tidal streams \cite{bovy2015galpy}.

For more details about how \verb+galpy+ works and what it is capable of refer to \citet{bovy2015galpy} and its corresponding documentation \cite{GALPY}. 

\subsection{Fitting process and output}\label{sec:GalRotpy-Fitting}

The \textbf{GalRotpy} visual interface is a great help when the behavior of rotation curves wants to be understood  or  explained, but even though the visual inspection of the different potentials allows the user to choose the best candidates to reproduce the data, and a set of parameters which at first glance seems to be correct, it is essential to find a better estimation of the parameters with their uncertainties.

To solve this problem we make use of the package \verb+emcee+ \cite{EMCEE, foreman2013emcee} which implements a particular Markov chain Monte Carlo (MCMC) algorithm proposed by \citet{goodman2010ensemble}, where instead of getting a best fit curve like a frequentist approach does, MCMC obtains the posterior probability distribution $P(\mathbf{\theta}\vert\mathcal{D}, \mathcal{M})$ with $\mathbf{\theta}$, $\mathcal{D}$ and $\mathcal{M}$ being the parameters involded, the data used and the model (composed rotation curve in our case) respectively, such that
\begin{equation}\label{eq:posterior}
P(\mathbf{\theta}\vert\mathcal{D}, \mathcal{M})=\frac{P(\mathcal{D}\vert \mathbf{\theta}, \mathcal{M})P(\mathbf{\theta}\vert\mathcal{M})}{P(\mathcal{D}\vert \mathcal{M})},
\end{equation}
where $\mathcal{L}:=P(\mathcal{D}\vert \mathbf{\theta}, \mathcal{M})$ is the likelihood, $\Pi:=P(\mathbf{\theta}\vert\mathcal{M})$ is the prior, and $Z:=P(\mathcal{D}\vert \mathcal{M})$ is the evidence (also known as marginal likelihood or normalization factor). Nevertheless, since the evidence is not considered in MCMC algorithms, we only need to input the likelihood and the prior distributions: for the likelihood we use a Gaussian distribution
\begin{equation}\label{eq:likelihood}
\mathcal{L}\propto\exp{\left(-\frac{1}{2}\sum_{i=1}^{N}\left[\frac{v_i^{data}-v_i^{model}}{v_i^{error}}\right]^2\right)},
\end{equation}
with $N$ being the number of data points, and for the prior we use a step-like distribution given by
\begin{equation}\label{eq:prior}
\Pi=\begin{cases}
\begin{array}{c}
1\\
0
\end{array} & \begin{array}{c}
if\;\theta>0\\
if\;\theta\leq0
\end{array}\end{cases},
\end{equation}
where $\theta$ refers to all parameters. It is worth noting that we chose this simple distributions in order to not to give several constrains during the fitting process since we lack of information of how the different contributions to the composed circular velocity will adapt to the given data. In case, there is previous knowledge which helps to constrain the parameters, it may be introduced into the probability distributions. For a further discussion about how to implement \verb+emcee+ refer to its documentation \cite{EMCEE}.

Now, with respect to how \textbf{GalRotpy} works, for each distribution we allow to their parameters to evolve only in a constrained way, due to the probability distribution given above; except for the bulge which can be modeled by means of two distributions: a spherical symmetric distribution or spheroidal distribution. The first one is described by means  of the Plummer potential, which is achieved leaving $a=0kpc$ as initial guess such that this parameter will not be used through the fitting process. However, if the user let $a>0kpc$ as initial guess, the bulge will be modeled by meas of the Miyamoto-Nagai potential.

Therefore, once the potentials and also the parameters' values, which seems to reproduce the data, have been chosen, the user has to click on the start button which closes the panel and will start the fitting process on the shell, where it is shown the dimension of the system (number of parameters consider in total) and it is asked to introduce the number o walkers (Markov chains) to be used; which must be an even number at least, twice the dimension. After that, the number of steps the user wants the walkers to take, has to be introduced in order to explore the parameters space. 

An initial set of parameters or initial guess is needed for \verb+emcee+ to work, thus, such set corresponds to those parameters approximated through the visual fitting and are saved in a file named \verb+init_guess_params.txt+. This initial guess needs to reproduce the data as well as possible, othervise the results are more likely to diverge.

The success of the fitting process in linked to how well the walkers behave, such that it is advised to take a big enough number of steps for a small number of walker (at least twice the dimension of the system) so, the user can verify what combination of components are the best for the rotation curve to be studied, and also can verify if the walkers actually converge, and if they do, the user can check whether they converge to physical values or not. Therefore, once one is sure which components to use and how well the walkers behave, in order to improve the estimation of the parameters, the fitting process can be run several times, each one using the very same input number of walkers and steps, making a new initial guess each time from half of the steps, thus the system evolves smoothly. If the user wants to run the process more than once, in the last run the walkers will take three times the number of input steps in order to have a better visualization of the walkers' behavior. 

It is worth noting that not for every possible combination of contributions \textbf{GalRotpy} will provided reliable results, since for some combination the walkers will diverge or converge to unphysical values. This problem may be addressed using a careful number of steps (small compared to the number of steps by which the walkers start to diverge) and running \textbf{GalRotpy} as many times as considered correct, nevertheless this procedure most of the times yields to nonphysical results, so it has to be applied carefully.

\begin{figure}[]
	\centering
	\hspace*{-0.6cm} 
	\includegraphics[width=1.13\columnwidth]{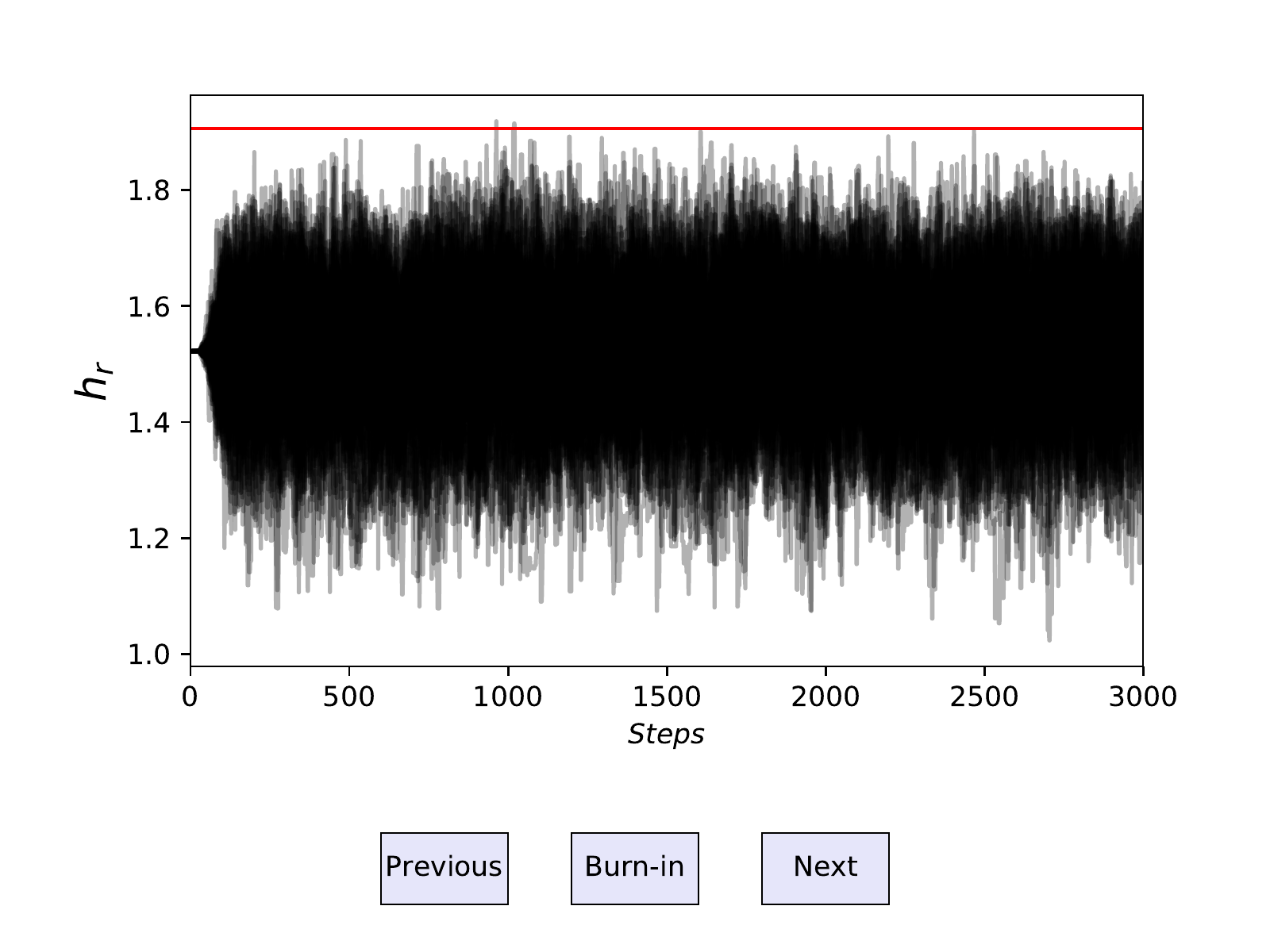}
	\caption{Panel to explore walkers' behavior for each parameter being considered. This particular example presents the behavior of $h_r$ for the fitting process shown in Fig. \ref{fig:m33} (top). The red line represents the corresponding initial guess value obtained by means of the visual inspection (fit).\\}
	\label{fig:walkers}
\end{figure}

After the walkers explore the parameters' space (fitting process), a window opens. Such window shows the walkers behavior (Markov chains) as it is shown in Fig. \ref{fig:walkers}. This window has three buttons: two of them allow the user to see the samples for each parameter being studied so, that it is easy to determine from which step the chains are actually converging. It means that it is possible to get rid of those steps which are not useful; for example in Fig. \ref{fig:walkers} the fact of getting rid of the first 500 steps, gives excellent results. Hence, when the user decides how many steps to burn in, after clicking on the named button, the window closes and the number of steps to be burn in, has to be introduced in the shell. 

Finally, this leads to three files: the first one is a text file named \verb+final_params.txt+ which includes the parameters' values obtained with their corresponding uncertainties for the $68\%$ and $95\%$ quantiles, and two plots: one shows the curve obtained with each of the contributions used, and the other one shows the credibility regions which are plotted using the package \verb+corner+ \cite{CORNER,corner2016}. In regards to the credibility, as can be seen in Fig. \ref{fig:NGC6361} and Fig. \ref{fig:m33} we have that, the inner dark region correspods to the $68\%$ likelihood, followed by fainter regio corresponding to the $95\%$ likelihood. The outer dark dotted region corresponds to the data beyond the $95\%$ likelihood.  

In case that the exponential disc is selected, the text file \verb+final_params.txt+ will also include the total mass of the given disc $M_d$, likewise, for both dark halos it will also be included the concentration parameter $X$ and the halo's total mass $M_h$ for a given cosmological overdensity $\Delta_c$, whose value is asked after \textbf{GalRotpy} panel is closed. Since these quantities are not included directly along the fitting process, they will not appear in the credibility regions plot.

\section{Results using GalRotpy}\label{sec:results}

For the purpose of showing how \textbf{GalRotpy} works, we use the disc galaxies M33 and NGC6361 as test cases to find a dynamical model that describes approximately the gravitational potential of the given galaxies. For M33 we are able to compare our results with those reported by \citet{lopez2017radial}.

\subsection{NGC6361 test case}

\begin{figure}[]
	\centering
	\includegraphics[scale=0.5]{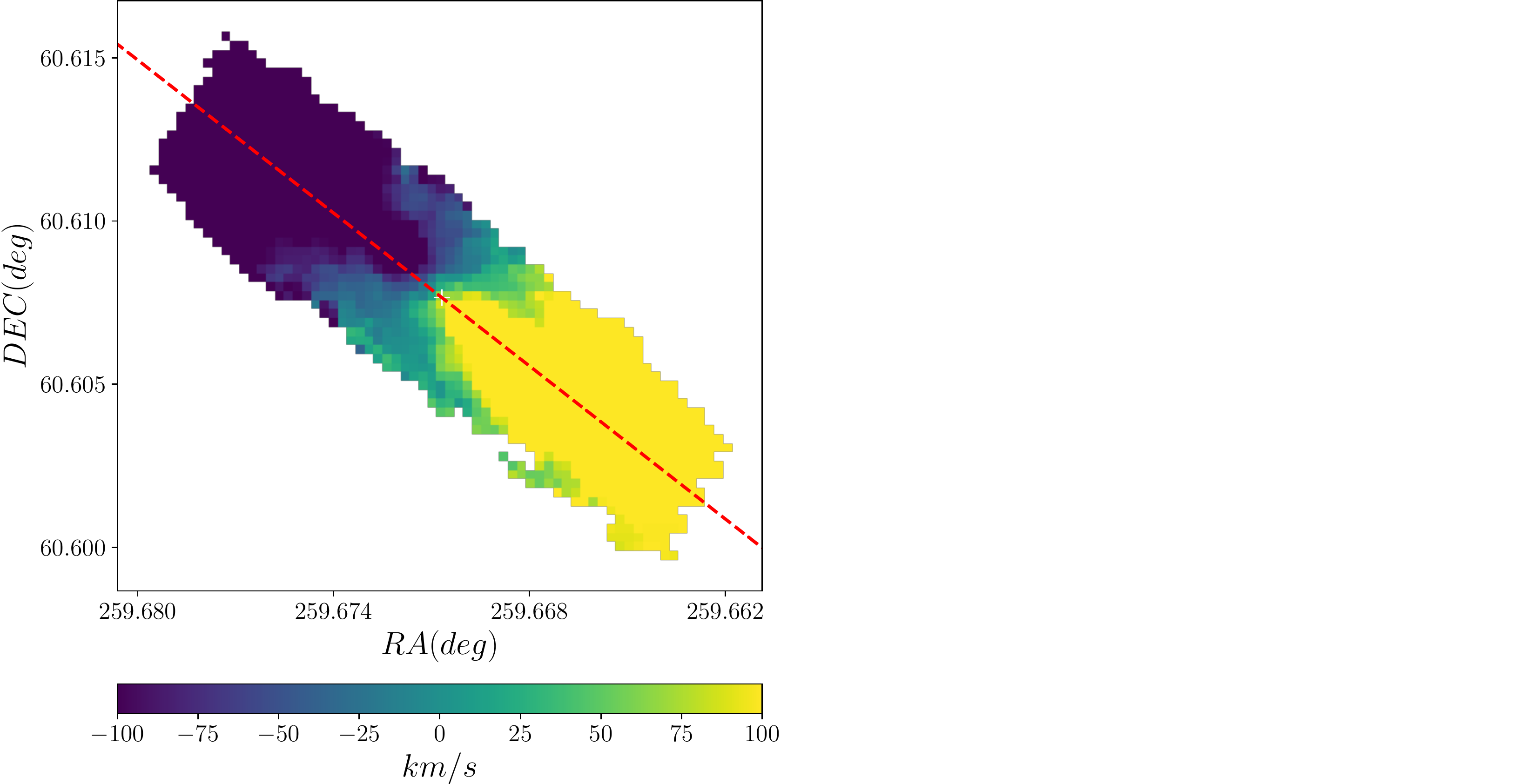}
	\caption{Gas velocity field of NGC6361 for the data product, provided by CALIFA collaboration. X and Y axis indicates right ascension ($RA$) and declination ($DEC$) respectively. The red line represents the major axis of the system.}
	\label{fig:Fi3}
\end{figure}

To get the rotation curve of NGC6361, we first make a selection of some galaxies from CALIFA (Calar Alto Legacy Integral Field Area) survey, which provides data cubes of more than 600 galaxies in the local universe with 0.005 $<$ z $<$ 0.03. CALIFA survey uses Integral Field Spectroscopy (IFS) to integrate the properties of images and spectroscopy. The data cubes have information about kinematic properties from emission and absorption lines, stellar populations, and other physical features of each galaxy in CALIFA survey sample \citep{garcia2015califa}. Then, we select the NGC6361 galaxy which is a spiral galaxy type (SAb edge-on) \cite{simbadngc6361} that does not present a bar-like structure in it. After that, we obtain from CALIFA survey the data product of NGC6361, one derived using PIPE3D, a technique implemented by \citet{sanchez2016pipe3d}. Based on the datacube of NGC6361 and the velocity map for \textsc{H}$\alpha$ emission line provided by CALIFA collaboration, we get the Fig. \ref{fig:Fi3}.

\begin{figure}[]
	\centering
	\includegraphics[scale=0.84]{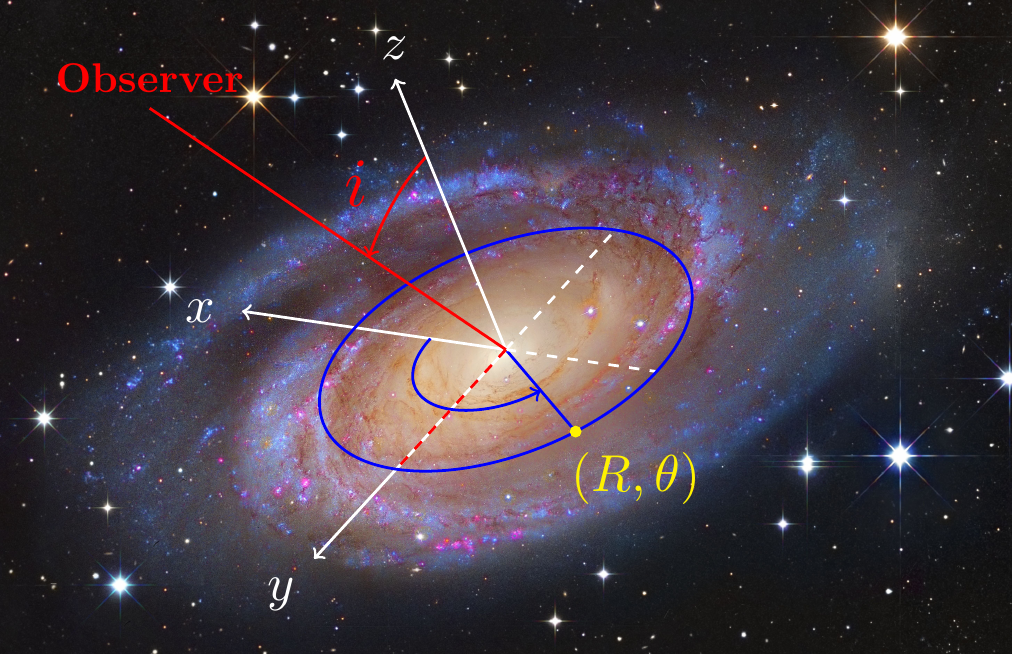}
	\includegraphics[scale=0.80]{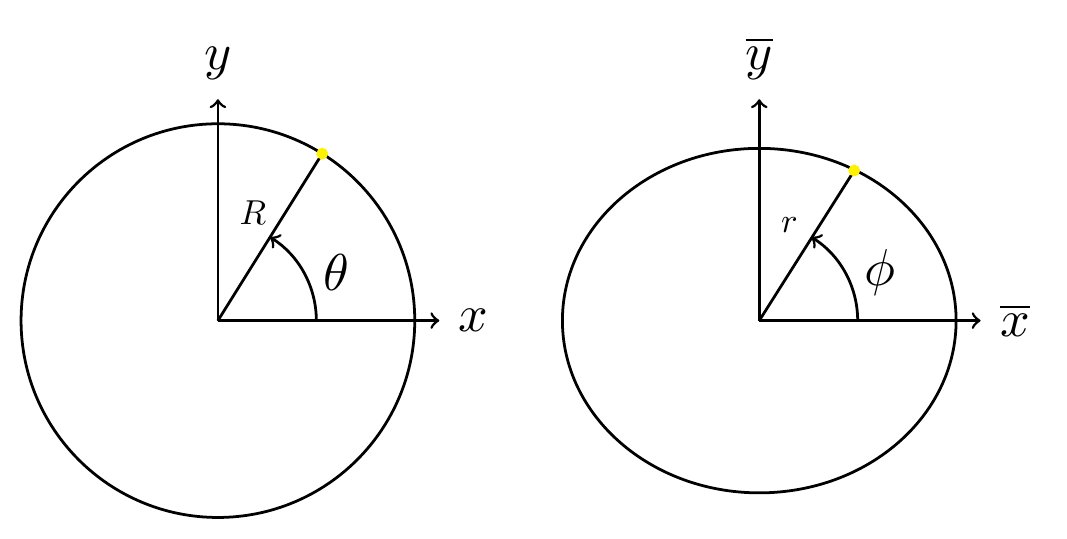}
	
	\caption{Set up used to derive the rotation curve from observations. At the top we have the trajectory (blue curve) followed by a star (yellow point) with polar coordinates $(R, \theta)$, with respect to the galaxy's coordinate system. We can see that the galaxy is inclined with respect to the observer's line of sight (red) by an angle $i$. At the bottom we have the representation of a trajectory as it is seen in the galaxy's plane (left) and in the observer's plane (right) with their respective polar coordinates, where it has been considered that the axis $x$ and $\overline{x}$ are parallel along the major axis. The galaxy used in this illustration (top) corresponds to the spiral galaxy M81, whose credits are\cite{nasa}:\\	
	\textbf{Image Credit}: Subaru Telescope, Hubble Space Telescope.\\
	\textbf{Processing and Copyright}: Roberto Colombari and Robert Gendler.\\}
	\label{fig:Fi4}
	
\end{figure}

In this example, the rotation curve of NGC6361 is obtained from the points over the kinematic center, taken over the radial coordinate on the gas velocity field, along the given major axis (see Fig. \ref{fig:Fi3}). To accomplish this task, we have defined two coordinate systems, each one with respect to a given plane: one of them is perpendicular to the line of sight, while the other is perpendicular to the galaxy's polar axis, such that the inclination angle between these planes is named $i$; see Fig. \ref{fig:Fi4} (top). Thus, if we set these coordinate systems in such a way that the axis $x$ and $\overline{x}$  are parallel, and coincide with the galaxy's major axis as seen by the observer, we can relate the position of a star over the galaxy's plane and the observer's plane, as it is illustrated in Fig. \ref{fig:Fi4} (bottom).

Consider a point (star) in the galaxy's plane, with position $(R, \theta)$ (see Fig. \ref{fig:Fi4} bottom-left),  and whose velocity in this coordinate basis is
\[
\textbf{V}_{\star}=\left(V_R\cos(\theta)-V_{\theta}\sin(\theta)\right)\hat{x}+\left(V_R\sin(\theta)+V_{\theta}\cos(\theta)\right)\hat{y}.
\] 
Now, the observer is capable of measuring only the component of the velocity along its line of sight i.e, along $\overline{z}$, which is given by $\hat{\overline{z}}\cdot\textbf{V}_{\star}$. Therefore, taking into account that from our set up the relations  $\hat{x}=\hat{\overline{x}}$, and $\hat{y}=\cos(i)\hat{\overline{y}}+\sin(i)\hat{\overline{z}}$, are satisfied, the velocity measured by the observer turns out to be\cite{beckman2004kinematic}
\begin{equation}
	\label{eq: observed velocity}
	V_{obs} = V_{sys}+V_\theta sin (i) cos (\theta)+V_R sin (i) sin(\theta).
\end{equation}

Here, an additional term named systemic velocity ($V_{sys}$) is added, which corresponds to the velocity of the galaxy as a whole (given by the spectroscopic redshift), while $V_R$ and $V_{\theta}$ represent the velocity along  the radial and tangential direction respectively; with $V_{\theta}$ being the velocity component we are interested in. 
For this situation is commonly assumed that $V_R$ can be neglected, and also, for simplicity only the velocities along the major axis are considered ($\theta=0$). Then, from (\ref{eq: observed velocity}) the circular velocity reads
\begin{equation}
V_\theta = \frac{V_{obs}-V_{sys}}{sin(i)}.
\end{equation}

In this case, we have assumed that the gas velocity follows approximately the galaxy potential like the stars velocity field.

\begin{table}[]
	\centering
	\caption{Set of parameters obtained using \textbf{GalRotpy} with their corresponding uncertainties for three different models for the rotation curve of NGC6361. For each dark halo distribution, the halo's total mass $M_h$ and the concentration parameter $X$ are given for $\Delta_c=97.2$.\\}
	\label{NGC6361_tab}
	\begin{tabular}{lclcc}
		\hline
		\multicolumn{5}{c}{\textbf{Model I}}                      \\ \hline
		\multicolumn{1}{c}{
			\textbf{Component}} & 
		\textbf{Parameter}  & 
		\textbf{Fit}        & 
		\textbf{68\%} 		& 
		\textbf{95\%} 
		\\ \hline
		NFW-Halo                              
		& \begin{tabular}[c]{@{}c@{}}
			$a(kpc)$\\
			$M_{0}(\times 10^{11}M_{\odot})$\\
			$\rho_{0}(\times 10^{8}M_{\odot}/kpc^3)$\\
			$X(\times 10)$\\ 
			$M_{h}(\times 10^{11}M_{\odot})$
		\end{tabular}            
		&\begin{tabular}[c]{@{}c@{}}
			$4.93$\\ 
			$3.04$\\
			$2.02$\\
			$5.27$\\
			$9.14$
		\end{tabular}            
		&\begin{tabular}[c]{@{}c@{}}
			$^{+0.21}_{-0.19}$\\ 
			$^{+0.16}_{-0.15}$\\
			$^{+0.15}_{-0.14}$\\
			$^{+0.14}_{-0.14}$\\
			$^{+0.40}_{-0.38}$
		\end{tabular}           
		&\begin{tabular}[c]{@{}c@{}}
			$^{+0.42}_{-0.37}$\\
			$^{+0.33}_{-0.28}$\\
			$^{+0.31}_{-0.27}$\\
			$^{+0.28}_{-0.27}$\\
			$^{+0.82}_{-0.72}$
		\end{tabular} 
		\\ \hline
		\multicolumn{5}{c}{\textbf{Model II}}                      \\ \hline
		\multicolumn{1}{c}{
			\textbf{Component}} & 
		\textbf{Parameter}  & 
		\textbf{Fit}        & 
		\textbf{68\%} 		& 
		\textbf{95\%} 
		\\ \hline
		Exponential Disc\;\;\;           & 
		\begin{tabular}[c]{@{}c@{}}
			$h_r(kpc)$\\
			$\Sigma_0(\times 10^{3}M_{\odot}/pc^{2})\;\;$\\ $M_{\star}(\times 10^{10}M_{\odot})$
		\end{tabular}    
		&\begin{tabular}[c]{@{}c@{}}
			$1.05$\\ 
			$2.84$\\ 
			$1.97$
		\end{tabular}         
		&\begin{tabular}[c]{@{}c@{}}
			$^{+0.08}_{-0.08}$\\
			$^{+0.11}_{-0.11}$\\
			$^{+0.26}_{-0.27}$
		\end{tabular}
		&\begin{tabular}[c]{@{}c@{}}
			$^{+0.15}_{-0.16}$\\ 
			$^{+0.22}_{-0.23}$\\ 
			$^{+0.49}_{-0.55}$
		\end{tabular}    
		\\ \hline
		Burkert-Halo\;\;                           & \begin{tabular}[c]{@{}c@{}}
			$a(kpc)$\\ 
			$\rho_0(\times 10^{8}M_{\odot}/kpc^3)$\\
			$X(\times 10)$\\
			$M_{h}(\times 10^{12}M_{\odot})$
		\end{tabular}
		&\begin{tabular}[c]{@{}c@{}}
			$6.79$\\ 
			$1.24$\\ 
			$4.51$\\ 
			$1.49$
		\end{tabular}         
		&\begin{tabular}[c]{@{}c@{}}
			$^{+1.25}_{-0.96}$\\ 
			$^{+0.32}_{-0.25}$\\
			$^{+0.37}_{-0.40}$\\
			$^{+0.43}_{-0.28}$ 
		\end{tabular}
		&\begin{tabular}[c]{@{}c@{}}
			$^{+2.91}_{-1.69}$\\
			$^{+0.69}_{-0.45}$\\ 
			$^{+0.81}_{-0.70}$\\
			$^{+1.14}_{-0.46}$
		\end{tabular} 
		\\ \hline
		\multicolumn{5}{c}{\textbf{Model III}}                     \\ \hline
		\multicolumn{1}{c}{
			\textbf{Component}} 	& 
		\textbf{Parameter}  	& 
		\textbf{Fit} 			& 
		\textbf{68\%} 			& 
		\textbf{95\%} 
		\\ \hline
		Thin Disc               
		& \begin{tabular}[c]{@{}c@{}}
			$a(\times 10^{-1}kpc)$\\ 
			$b(\times 10^{-1}kpc)$\\
			$M_{\star}(\times 10^{10}M_{\odot})$
		\end{tabular}    
		&\begin{tabular}[c]{@{}c@{}}
			$6.56$\\ 
			$6.64$\\ 
			$2.42$
		\end{tabular}        
		&\begin{tabular}[c]{@{}c@{}}
			$^{+4.48}_{-4.46}$\\ 
			$^{+4.46}_{-4.49}$\\
			$^{+0.28}_{-0.28}$
		\end{tabular}
		&\begin{tabular}[c]{@{}c@{}}
			$^{+6.58}_{-6.22}$\\ 
			$^{+6.58}_{-6.30}$\\ 
			$^{+0.59}_{-0.55}$
		\end{tabular}   
		\\ \hline
		Burkert-Halo\;\;                           & \begin{tabular}[c]{@{}c@{}}
			$a(kpc)$\\ 
			$\rho_0(\times 10^{8}M_{\odot}/kpc^3)$\\
			$X(\times 10)$\\
			$M_{h}(\times 10^{12}M_{\odot})$
		\end{tabular}
		&\begin{tabular}[c]{@{}c@{}}
			$5.70$\\ 
			$1.53$\\ 
			$4.87$\\
			$1.11$
		\end{tabular}         
		&\begin{tabular}[c]{@{}c@{}}
			$^{+0.73}_{-0.61}$\\ 
			$^{+0.31}_{-0.26}$\\
			$^{+0.35}_{-0.33}$\\
			$^{+0.19}_{-0.15}$ 
		\end{tabular}
		&\begin{tabular}[c]{@{}c@{}}
			$^{+1.62}_{-1.12}$\\
			$^{+0.68}_{-0.47}$\\
			$^{+0.71}_{-0.63}$\\ 
			$^{+0.46}_{-0.25}$
		\end{tabular} 
		\\ \hline
	\end{tabular}
\end{table}

\begin{figure*}[]
	\centering
	\includegraphics[scale=1]{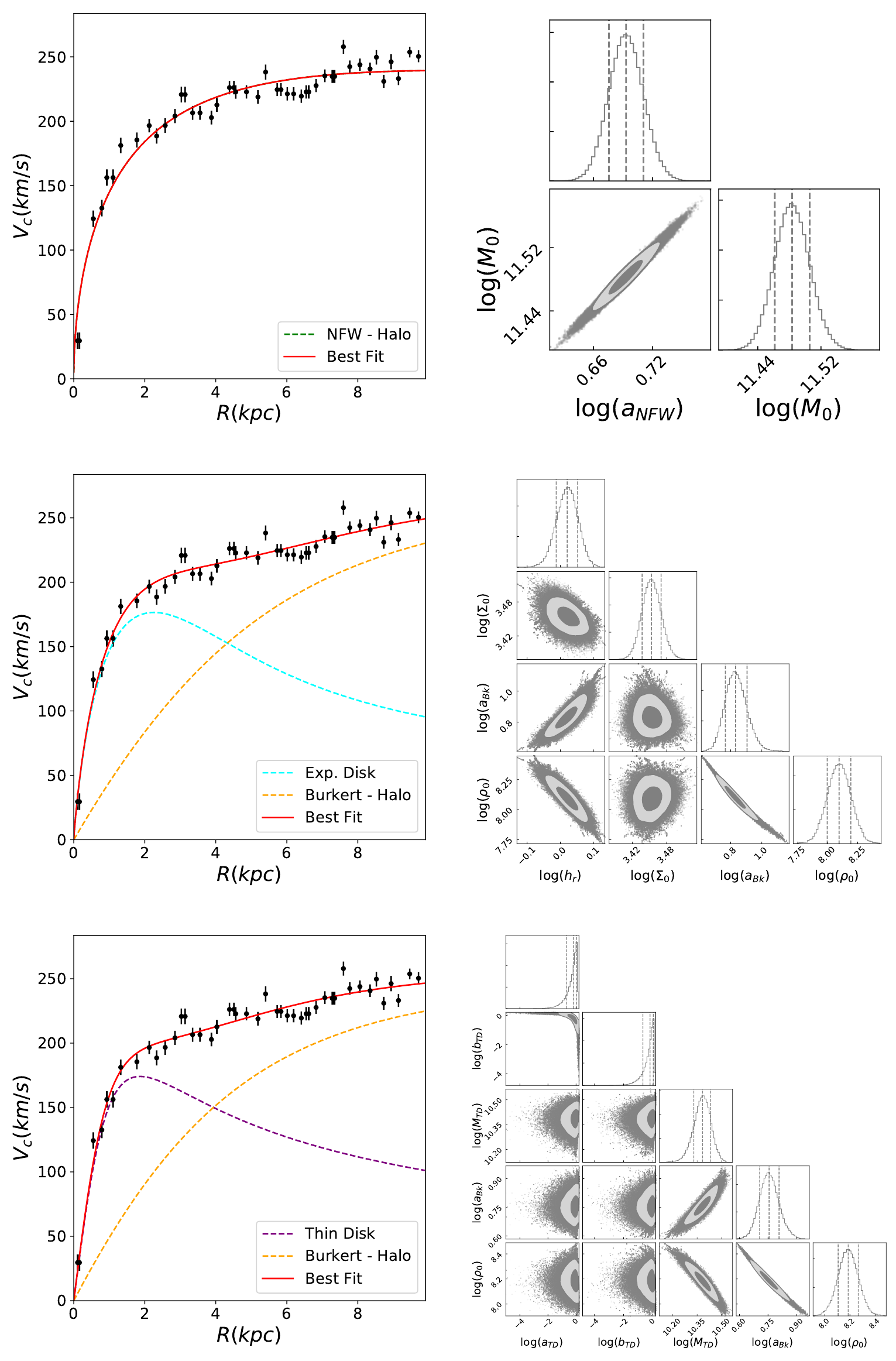}
	\caption{Rotation curve (left) and credibility regions (right) for three different models which attempt to reproduce the rotation curve of the disc galaxy NGC6361. The models are: NFW dark halo (top), Burkert dark halo + Exponential disc (middle) and  Burkert dark halo + Thin disc (bottom).}
	\label{fig:NGC6361}
\end{figure*}

At this point we have described how it is possible to obtain the rotation curve from observations, therefore, from  now on we will focus on the analysis of this curve using \textbf{GalRotpy}.

In the literature is often found that for a given value of  $\Delta_c$, dark halos contributions are parametrized directly through its total mass $M_h$ and the concentration parameter $X$; mostly the NFW profile. However, although \textbf{GalRotpy} does not use this parameters to fit the rotation curve, they can be derived as it is discussed in section  \ref{sec:DH mass}, while  the total mass $M_{\star}$ corresponding to the exponential disc is easily calculated by using (\ref{eq:mass exp disc}).

Therefore, from its rotation curve, now it is possible to characterize NGC6361. Here, we show three models which provide reliable parameters' values (see Table \ref{NGC6361_tab}), where we run the fit process twice for 100 walkers, 3000 steps and  $\Delta_c=97.2$.

We modeled this galaxy's rotation curve using both dark matter halo profiles available in \textbf{GalRotpy}. For the NFW profile we found that reliable results  are obtained only when this profile is applied, whose rotation curve is shown in Fig. \ref{fig:NGC6361} (top). It means that this profile is capable of reproducing the rotation curve by itself, which suggests that within the corresponding uncertainties, NGC6361 is a galaxy dominated by dark matter, for a halo with a mass $M_h\sim 10^{12}M_{\odot}$ concentrated within a radius $r_c\sim 10^2kpc$. On the other hand, for the Burkert profile, we found that it is not capable of reproducing the given rotation curve by itself, which is expected considering its behavior (see Fig. \ref{fig:Fig_BK_parameters}), since this profile cannot reproduce the cusp in the inner region. We obtained two models compatible with the data, each one for a different disc profile: an exponential disc (Fig. \ref{fig:NGC6361} - middle) and a thin disc (Fig. \ref{fig:NGC6361} - bottom). Such discs structures are dominant in the inner regions, approximately $R<4kpc$, beyond this limit in both cases the disc contribution starts decreasing rapidly and the dark halo is the dominant dynamical component . For this profile, we also found that within the corresponding uncertainties the dark halo is dominant with a mass $M_h\sim 10^{12} M_{\odot}$ enclosed within a radius $r_c\sim 10^2kpc$ with respect to a mass $M_{\star}\sim 10^{10}M_{\odot}$ for the disc contribution.

\subsection{M33 test case}

\begin{table}[]
	\centering
	\caption{Set of parameters obtained using \textbf{GalRotpy} with their corresponding uncertainties for two models for the rotation curve of M33. For each dark halo distribution the halo's total mass $M_h$ and the concentration parameter $X$ are given for $\Delta_c=97.2$.\\}
	\label{m33_gal}
	\begin{tabular}{lclcc}
		\hline
		\multicolumn{5}{c}{\textbf{Model I}}                      \\ \hline
		\multicolumn{1}{c}{
			\textbf{Component}} & 
		\textbf{Parameter}  & 
		\textbf{Fit}        & 
		\textbf{68\%} 		& 
		\textbf{95\%} 
		\\ \hline
		Exponential Disc\;\;\;           & 
		\begin{tabular}[c]{@{}c@{}}
			$h_r(kpc)$\\
			$\Sigma_0(\times 10^{2}M_{\odot}/pc^{2})\;\;$\\ $M_{\star}(\times 10^{9}M_{\odot})$
		\end{tabular}    
		&\begin{tabular}[c]{@{}c@{}}
			$1.52$\\ 
			$2.50$\\ 
			$3.61$
		\end{tabular}         
		&\begin{tabular}[c]{@{}c@{}}
			$^{+0.10}_{-0.11}$\\
			$^{+0.37}_{-0.43}$\\
			$^{+0.96}_{-0.91}$
		\end{tabular}
		&\begin{tabular}[c]{@{}c@{}}
			$^{+0.20}_{-0.23}$\\ 
			$^{+0.66}_{-0.90}$\\ 
			$^{+1.89}_{-1.74}$
		\end{tabular}    
		\\ \hline
		NFW-Halo                              
		& \begin{tabular}[c]{@{}c@{}}
			$a(\times 10 kpc)$\\
			$M_{0}(\times 10^{11}M_{\odot})$\\
			$\rho_0(\times 10^{6}M_{\odot}/kpc^3)$\\ 
			$X(\times 10)$\\
			$M_{h}(\times 10^{11}M_{\odot})$		
		\end{tabular}            
		&\begin{tabular}[c]{@{}c@{}}
			$1.46$\\ 
			$2.37$\\
			$6.05$\\
			$1.37$\\
			$4.16$
		\end{tabular}            
		&\begin{tabular}[c]{@{}c@{}}
			$^{+0.42}_{-0.29}$\\ 
			$^{+0.91}_{-0.55}$\\
			$^{+2.96}_{-2.13}$\\
			$^{+0.24}_{-0.22}$\\
			$^{+1.11}_{-0.72}$
		\end{tabular}           
		&\begin{tabular}[c]{@{}c@{}}
			$^{+1.02}_{-0.49}$\\
			$^{+2.45}_{-0.91}$\\
			$^{+6.88}_{-3.55}$\\
			$^{+0.48}_{-0.42}$\\
			$^{+2.86}_{-1.21}$
		\end{tabular} 
		\\ \hline
		\multicolumn{5}{c}{\textbf{Model II}}                     \\ \hline
		\multicolumn{1}{c}{
			\textbf{Component}} 	& 
		\textbf{Parameter}  	& 
		\textbf{Fit} 			& 
		\textbf{68\%} 			& 
		\textbf{95\%} 
		\\ \hline
		Exponential Disc               
		& \begin{tabular}[c]{@{}c@{}}
			$h_r(kpc)$\\ 
			$\Sigma_0(\times 10^{2}M_{\odot}/pc^{2})\;\;$\\ $M_{\star}(\times 10^{9}M_{\odot})$
		\end{tabular}    
		&\begin{tabular}[c]{@{}c@{}}
			$1.35$\\ 
			$4.74$\\ 
			$5.47$
		\end{tabular}        
		&\begin{tabular}[c]{@{}c@{}}
			$^{+0.11}_{-0.11}$\\ 
			$^{+0.15}_{-0.15}$\\
			$^{+0.88}_{-0.82}$
		\end{tabular}
		&\begin{tabular}[c]{@{}c@{}}
			$^{+0.21}_{-0.21}$\\ 
			$^{+0.31}_{-0.30}$\\ 
			$^{+1.71}_{-1.52}$
		\end{tabular}   
		\\ \hline
		Burkert-Halo\;\;                           & \begin{tabular}[c]{@{}c@{}}
			$a(kpc)$\\ 
			$\rho_0(\times 10^{7}M_{\odot}/kpc^3)$\\
			$X(\times 10)$\\
			$M_{h}(\times 10^{11}M_{\odot})$
		\end{tabular}
		&\begin{tabular}[c]{@{}c@{}}
			$6.61$\\ 
			$2.66$\\
			$2.52$\\ 
			$2.39$
		\end{tabular}         
		&\begin{tabular}[c]{@{}c@{}}
			$^{+0.84}_{-0.71}$\\ 
			$^{+0.63}_{-0.52}$\\
			$^{+0.22}_{-0.20}$\\
			$^{+0.29}_{-0.23}$ 
		\end{tabular}
		&\begin{tabular}[c]{@{}c@{}}
			$^{+1.71}_{-1.27}$\\
			$^{+1.32}_{-0.89}$\\ 
			$^{+0.43}_{-0.37}$\\
			$^{+0.61}_{-0.40}$
		\end{tabular} 
		\\ \hline \\
	\end{tabular}
\end{table}

\begin{table}[]
	\caption{Set of parameters reported by \citet{lopez2017radial}. Here $\Delta_c=97.2$.\\}
	\label{m33_fune}
	\begin{tabular}{lclcc}
		\hline
		\multicolumn{5}{c}{\textbf{Model I}}                      \\ \hline
		\multicolumn{1}{c}{
			\textbf{Component}} & 
		\textbf{Parameter}  & 
		\textbf{Fit}        & 
		\textbf{Uncertainty} 		
		\\ \hline
		Stellar-Gas-Halo         & 
		\begin{tabular}[c]{@{}c@{}}
			$M_{\star}(\times 10^{9}M_{\odot})$
		\end{tabular}    
		&\begin{tabular}[c]{@{}c@{}}
			$4.9$
		\end{tabular}         
		&\begin{tabular}[c]{@{}c@{}}
			$\pm 1.5$
		\end{tabular}
		
		\\ \hline
		NFW-Halo                              
		& \begin{tabular}[c]{@{}c@{}}
			$M_{h}(\times 10^{11}M_{\odot})$\\
			$X$
		\end{tabular}            
		&\begin{tabular}[c]{@{}c@{}}
			$5.4$\\
			$9.5$
		\end{tabular}            
		&\begin{tabular}[c]{@{}c@{}}
			$\pm 0.6$\\
			$\pm.0.7$
		\end{tabular}           
		\\ \hline
		\multicolumn{5}{c}{\textbf{Model II}}                     \\ \hline
		\multicolumn{1}{c}{
			\textbf{Component}} 	& 
		\textbf{Parameter}  	& 
		\textbf{Fit} 			& 
		\textbf{Uncertainty} 			&  
		\\ \hline
		Stellar-Gas-Halo            & 
		\begin{tabular}[c]{@{}c@{}}
			$M_{\star}(\times 10^{9}M_{\odot})$
		\end{tabular}    
		&\begin{tabular}[c]{@{}c@{}}
			$\;4.9$
		\end{tabular}         
		&\begin{tabular}[c]{@{}c@{}}
			$\pm 1.5$
		\end{tabular}   
		\\ \hline
		Burkert-Halo\;\;                           & \begin{tabular}[c]{@{}c@{}}
			$a(kpc)$\\ 
			$\rho_0(\times 10^{6}M_{\odot}/kpc^3)$\\
			$M_h$
		\end{tabular}
		&\begin{tabular}[c]{@{}c@{}}
			$9.5$\\ 
			$12.3$\\ 
			$3.0$
		\end{tabular}         
		&\begin{tabular}[c]{@{}c@{}}
			$\pm 0.6$\\ 
			$\pm 1.0$\\
			$\pm 0.8$ 
		\end{tabular}
		
		\\ \hline 
	\end{tabular}
\end{table}

\begin{figure*}[htbp]
	\centering
	\includegraphics[scale=1]{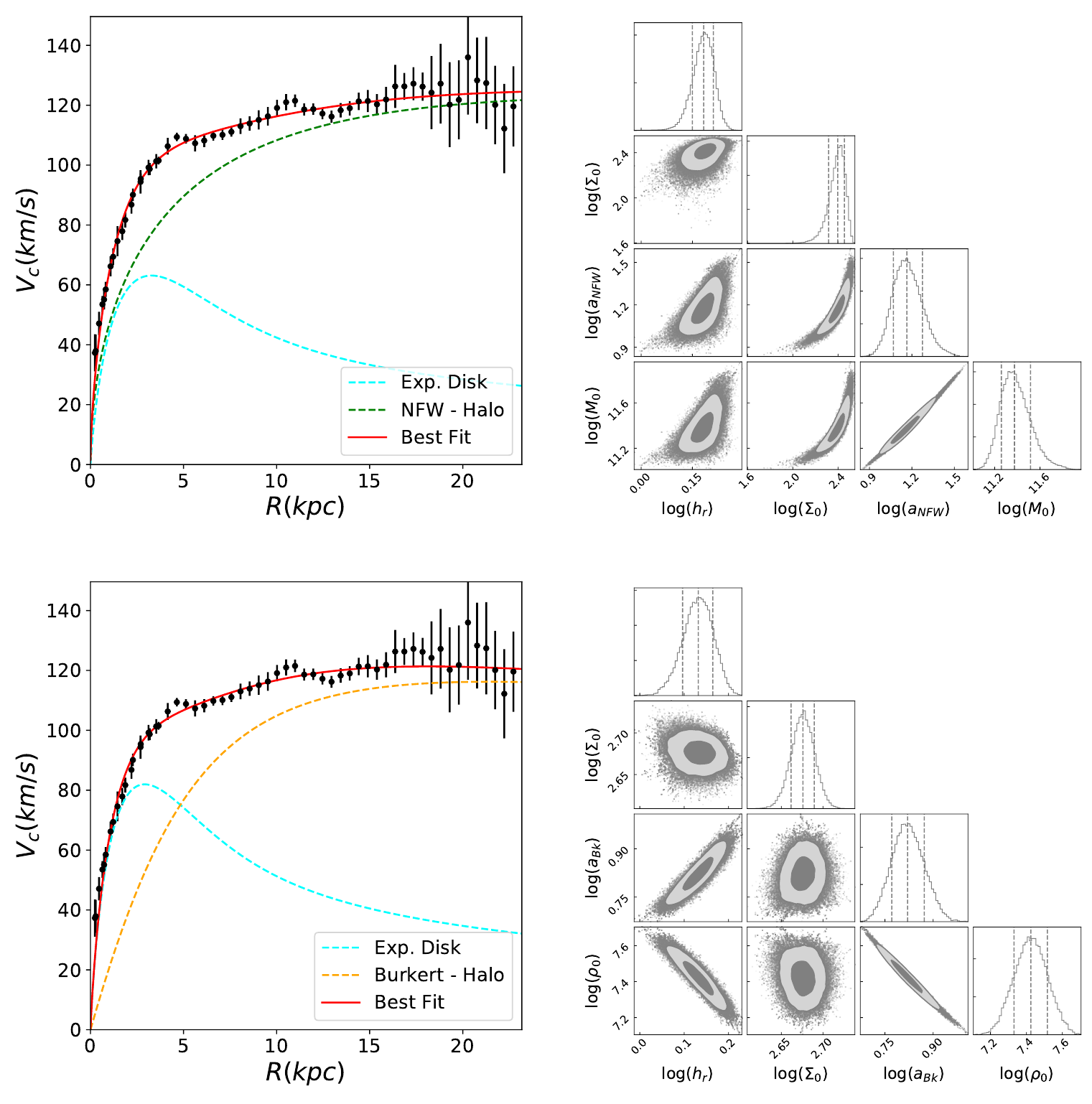}
	\caption{Rotation curve (left) and credibility regions (right) for two different models which attempt to reproduce the rotation curve of the disc galaxy M33. The models are: NFW dark halo + Exponential disc (top) and Burkert dark halo + Exponential disc (bottom). The rotation curve was taken from \citet{corbelli2014dynamical}.\\}
	\label{fig:m33}
\end{figure*}

We now focus on M33 which is a spiral galaxy type (SA(s)cd, Face -on) \cite{simbadm33} without a bar-like structure. We will characterize this galaxy based on the rotation curve taken from \citet{corbelli2014dynamical}, such that following \citet{lopez2017radial}, we have two models: in both cases we use an exponential disc potential to model the stellar and gaseous contribution, while the dark halo is modeled through NFW and Burkert profiles. 

Here, for both models we have a system of six dimensions (parameters) where we run the fit process twice for 100 walkers, 1000 steps and $\Delta_c=97.2$\cite{lopez2017radial}. Each parameter being considered converge as shown in Fig. \ref{fig:walkers}, such that the models are well adapted to the data. The parameters obtained are presented in Table \ref{m33_gal}, which within the corresponding uncertainties agree with those reported in \citet{lopez2017radial}, presented in Table \ref{m33_fune}.

From Fig. \ref{fig:m33} we can see that beyond $R\approx 5kpc$ for both models, the dark halo is the dominant contribution, and it rules the dynamics of the stars at this radius. However, the NFW profile presents a more significant contribution for the dynamics in the inner region, producing a more massive halo than the Burkert profile in Model II. On the other hand, since the Burkert profile is not capable of reproducing the cusp presented in the inner region ($R<5kpc$), thus, to compensate this fact the exponential disc turns out to be more massive than it is in Model I, then the baryonic matter is dominant in the inner region. 

\section{Conclusions}\label{sec:conclusions}

In this paper we have presented \textbf{GalRotpy}, which is a tool for the real-time composition of rotation curves of disc-like galaxies, being a straightforward and powerful method to study the behavior of rotation curves. This method gives an approximation to the dynamics of stellar systems and its global gravitational features. Thus, \textbf{GalRotpy} allows the user to check the presence of an assumed mass type component in an observed rotation curve, by including or removing a mass model quickly, then by means of a MCMC parametric fit, it is possible to verify if in fact, the contributions chosen are compatible with the data. From this fit process \textbf{GalRotpy} provides an estimation of the parameters involved with their uncertainties within the $68\%$ and $95\%$ likelihood, along  with a plot of the credibility regions associated to the intrinsic parameters of each contribution applied (those associated to the different profiles), by which it is easy to infer the main mass contribution quantitatively, in a galaxy from the mass ratios between pairs of mass components. Especially the bulge to disc and, the disc to dark matter halo ratios are relevant. \textbf{GalRotpy} also provides a plot which includes the composed rotation curve and its corresponding contributions, so it is possible to study qualitatively the influence of each component to the dynamics of the stellar system.

In order to present the capabilities of \textbf{GalRotpy} we have performed the analysis for the disc galaxies NGC6361 and M33. For NGC6361 we found three models consistent with the data: when the dark halo was modeled with the NFW profile, we obtained that this profile by itself is capable of reproducing the rotation curve, suggesting that the dark halo is the dominant dynamical component for this galaxy. Nevertheless, when the dark halo was modeled  with the Burkert profile unlike the foregoing model, since the Burkert profile is not capable of reproducing the inner cusp of this rotation curve, the presence of an additional structure was essential. For this task, we added a disc structure using an exponential disc and also a thin disc, such that in either case we have that the dynamical behavior in the inner region (approximately  $R<4kpc$) is dominated by the disc structure. With respect to M33 we applied the models suggested in \citet{lopez2017radial}, where our results are qualitatively and quantitatively in agreement with what they have reported. 

\section{Acknowledgements}

We thank PhD Jorge Barrera-Ballesteros and PhD Sebasti\'an S\'anchez for their collaboration on generating the kinematic map of NGC6361 from CALIFA survey. In the same way, we thank PhD E. L\'opez Fune, PhD P. Salucci and PhD E. Corbelli by allow us to use the rotation curve data of M33 galaxy. We are grateful to PhD P. Salucci for their suggestions on using the Burkert profile for an optional, not $\Lambda$-CDM scenario giving broad capabilities to GalRotpy for building rotation curves of disc-like galaxies. Finally, we especially acknowledge to PhD Leonardo Casta\~neda for promoting the interest in this work conducting the lectures on Galactic Dynamics at Observatorio Astron\'omico Nacional.


\bibliographystyle{aipauth4-1} 
\bibliography{References}

\end{document}